\PassOptionsToPackage{table,usenames,dvipsnames}{xcolor}
\documentclass[sigconf]{aamas}

\usepackage{balance} 

\setcopyright{ifaamas}
\acmConference[AAMAS '24]{Full version of the paper appearing in Proc.\@ of the 23rd International Conference
on Autonomous Agents and Multiagent Systems (AAMAS 2024)}{May 6 -- 10, 2024}
{Auckland, New Zealand}{N.~Alechina, V.~Dignum, M.~Dastani, J.S.~Sichman (eds.)}
\copyrightyear{2024}
\acmYear{2024}
\acmDOI{}
\acmPrice{}
\acmISBN{}

\acmSubmissionID{862}

\usepackage{mathtools} 
\usepackage{amsthm} 
\usepackage{bm} 
\usepackage{appendix}
\usepackage{bold-extra}

\usepackage{multirow}
\usepackage{booktabs} 
\usepackage{array} 
\usepackage{graphicx}
\graphicspath{./figs}
\usepackage{grffile} 
\usepackage{caption}
\usepackage{subcaption} 
\usepackage{enumitem} 
\setlist[itemize]{leftmargin=15pt,labelsep=5pt,noitemsep,topsep=5pt}
\setlist[enumerate]{leftmargin=15pt,labelsep=5pt,noitemsep,topsep=5pt}
\newlist{inline}{enumerate*}{1}
\setlist[inline]{before=\unskip{: }, itemjoin={{; }}, itemjoin*={{; and }}, label={(\roman*)}}
\usepackage[ruled,linesnumbered,noend]{algorithm2e}

\newcommand{\posreals}{\mathbb{R}_{>0}}
\newcommand{\posints}{\mathbb{Z}_{>0}}
\newcommand{\nonnegreals}{\mathbb{R}_{\geq 0}}

\newtheorem{theorem}{Theorem}[section]{\bfseries}{\itshape} 
\newtheorem{lemma}[theorem]{Lemma}{\bfseries}{\itshape}
{\bfseries}{\itshape}
\theoremstyle{definition}
\newtheorem{remark}[theorem]{Remark} 
\newtheorem{proposition}[theorem]{Proposition} 
\newtheorem{corollary}[theorem]{Corollary} 
\newcounter{prob}
\newtheorem{problem}[prob]{Problem} 
\makeatletter

\newcommand{\Rmnum}[1]{\expandafter\@slowromancap\romannumeral #1@}
\makeatother

\newcommand{\valfun}{f}

\newcommand{\maxwelfare}{\textsc{MaxWelfare}}
\newcommand{\maxleximin}{\textsc{MaxLeximin}}
\newcommand{\maxwelfarefair}{\textsc{MaxWelfareFair}}

\newcommand{\maxtbmrm}{\mbox{\textsc{MaxTB-MRM}}}

\newcommand{\subsetcol}{\mathcal{L}}
\newcommand{\match}{\mathcal{M}}
\newcommand{\trade}{\mathcal{T}}
\newcommand{\totvalnotrade}{\sigma_0}
\newcommand{\totval}{\sigma}
\DeclareMathOperator{\welfare}{welfare}

\DeclareMathOperator{\weight}{weight}
\newcommand{\touchet}[1]{\texttt{Touchet#1}}
\newcommand{\yakima}[1]{\texttt{Yakima#1}}
\newcommand{\betahigh}{\beta_{h}}
\newcommand{\betalow}{\beta_{\ell}}

\newcommand{\QED}{\hfill\rule{2mm}{2mm}}

\newcommand{\para}[1]{\smallskip\noindent\textbf{#1.}}

\newcommand{\cnp}{\textbf{NP}}

\newcommand{\xthrc}{\mbox{X3C}}
\newcommand{\ssum}{\mbox{\textsc{Ssum}}}

\newcommand{\titlestr}{Value-based Resource Matching with Fairness Criteria: Application to Agricultural Water Trading}

\title[\titelstr]{\titlestr{}}

\renewcommand{\email}[1]{}

\author{Abhijin Adiga}
\authornote{Both authors contributed equally to this work.}
\affiliation{
  \institution{University of Virginia}
  \city{Charlottesville}
  \state{VA}
  \country{USA}}
  \email{abhijin@virginia.edu}

  \author{Yohai Trabelsi}
\authornotemark[1]
\affiliation{
  \institution{Bar-Ilan University}
  \city{Ramat Gan}
  \country{Israel}}
\email{yohai.trabelsi@gmail.com}

\author{Tanvir Ferdousi}
\affiliation{
  \institution{University of Virginia}
  \city{Charlottesville}
  \state{VA}
  \country{USA}}
  \email{tanvir@virginia.edu}

\author{Madhav Marathe}
\affiliation{
  \institution{University of Virginia}
  \city{Charlottesville}
  \state{VA}
  \country{USA}}
  \email{marathe@virginia.edu}

 \author{S. S. Ravi}
\affiliation{
  \institution{University of Virginia}
  \city{Charlottesville}
  \state{VA}
  \country{USA}}
  \email{ssravi0@gmail.com}

\author{Samarth Swarup}
\affiliation{
  \institution{University of Virginia}
  \city{Charlottesville}
  \state{VA}
  \country{USA}}
  \email{swarup@virginia.edu}

\author{Anil Kumar Vullikanti}
\affiliation{
  \institution{University of Virginia}
  \city{Charlottesville}
  \state{VA}
  \country{USA}}
  \email{vsakumar@virginia.edu}

  \author{Mandy L. Wilson}
\affiliation{
  \institution{University of Virginia}
  \city{Charlottesville}
  \state{VA}
  \country{USA}}
  \email{alw4ey@virginia.edu}

\author{Sarit Kraus}
\affiliation{
  \institution{Bar-Ilan University}
  \city{Ramat Gan}
  \country{Israel}}
\email{sarit@cs.biu.ac.il}

\author{Reetwika Basu}
\affiliation{
  \institution{Washington State University}
  \city{Pullman}
  \state{WA}
  \country{USA}}
  \email{reetwika.basu@wsu.edu}

\author{Supriya Savalkar}
\affiliation{
  \institution{Washington State University}
  \city{Pullman}
  \state{WA}
  \country{USA}}
  \email{supriya.savalkar@wsu.edu}

\author{Matthew Yourek}
\affiliation{
  \institution{Washington State University}
  \city{Pullman}
  \state{WA}
  \country{USA}}
  \email{matthew.yourek@wsu.edu}

\author{Michael Brady}
\affiliation{
  \institution{Washington State University}
  \city{Pullman}
  \state{WA}
  \country{USA}}
  \email{bradym@wsu.edu}

\author{Kirti Rajagopalan}
\affiliation{
  \institution{Washington State University}
  \city{Pullman}
  \state{WA}
  \country{USA}}
  \email{kirtir@wsu.edu}

\author{Jonathan Yoder}
\affiliation{
  \institution{Washington State University}
  \city{Pullman}
  \state{WA}
  \country{USA}}
  \email{yoder@wsu.edu}

\begin{abstract}
    Optimal allocation of agricultural water in the event of droughts is an important global problem. In addressing this problem, many aspects, including the welfare of farmers, the economy, and the environment, must be considered. Under this backdrop, our work focuses on several resource-matching problems accounting for agents with multi-crop portfolios, geographic constraints, and fairness. First, we address a matching problem where the goal is to maximize a welfare function in two-sided markets where buyers' requirements and sellers' supplies are represented by value functions that assign prices (or costs) to specified volumes of water. For the setting where the value functions satisfy certain monotonicity properties, we present an efficient algorithm that maximizes a social welfare function. When there are minimum water requirement constraints, we present a randomized algorithm which ensures that the constraints are satisfied in expectation. For a single seller--multiple buyers setting with fairness constraints, we design an efficient algorithm that maximizes the minimum level of satisfaction of any buyer. We also present computational complexity results that highlight the limits on the generalizability of our results. We evaluate the algorithms developed in our work with experiments on both real-world and synthetic data sets with respect to drought severity, value functions, and seniority of agents.

\end{abstract}


\newcommand{\BibTeX}{\rm B\kern-.05em{\sc i\kern-.025em b}\kern-.08em\TeX}

\begin{document}


\pagestyle{fancy}
\fancyhead{}


\maketitle 

\section{Introduction}
\label{sec:intro}

\subsection{Background}
The growth in the global population has led to a significant increase in
demand for agricultural and urban water supplies~\cite{fao2022}.  However, water supply augmentation has reached its
limit~\cite{chong2006water}.  Furthermore, climate change has led to an
increased occurrence of droughts, which, in turn, lead to severe water
shortages~\cite{kallis2008droughts}.  Water markets have been widely
proposed as an effective means of water reallocation during such
shortages~\cite{chong2006water}, and several formal and informal markets
have emerged across the
world~\cite{wheeler2021water,howe2013interbasin,howe2003water}. A widely
proposed (but much debated) approach is \emph{socially optimal water
allocation}, where water is transferred from low-value to high-value
agricultural
applications~\cite{chong2006water,howe2013interbasin,xu2018two}. Much of
the work in this regard has focused on elaborate modeling of agricultural,
hydrological and economic aspects of the problem as explored through
complex agent-based models~(see,
e.g.,~\cite{raffensperger2009deterministic,raffensperger2017,reetwika2023thesis,sharghi2022uncertainty}).
Some references have addressed computational aspects of such models (see,
e.g.,~\cite{liu2016optimizing,li2017stability}).

\begin{figure*}[htb]
\centering
\includegraphics[height=1.5in]{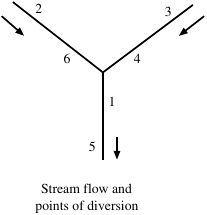}\hspace{2em} 
\includegraphics[height=1.5in]{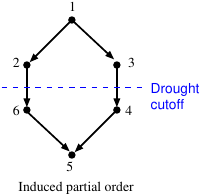}\hspace{2em} 
\includegraphics[height=1.5in]{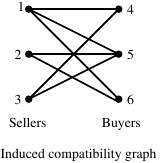}\hspace{2em} 
\includegraphics[height=1.5in]{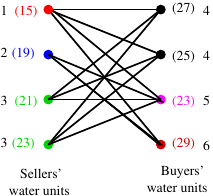}
\caption{Example. \normalfont{Going from left to right, the first panel
shows a sample stream flow and points of diversion for different trading
agents. In the second panel, each directed edge $(a,b)$ indicates that $a >
b$, that is, $a$ has higher priority than $b$. When there is no directed
edge between two nodes, the interpretation is that they belong to different
streams. The drought cutoff corresponds to the scenario where water is
available to agents~$1$,~$2$, and~$3$. In the compatibility graph shown in
the third panel, water units and their values have not yet been factored
in. The rightmost panel shows the corresponding resources--needs bipartite
graph along with the water units of the buyers and sellers and the
corresponding compatibility graph.  Here, Seller~3 has, and Buyer~4 needs,
two water units, while the others have or need only one water unit each.
The values of the water units are shown in parentheses.}}
\label{fig:example}
\end{figure*}

While our work is applicable to many water market settings, it is motivated
by the river water allocation mechanisms for agriculture in the western
US~\cite{brajer1989strengths,leonard2019expanding,booker1994modeling}.
Here, water is allocated according to the prior appropriation doctrine,
which induces a seniority ordering among
farmers~\cite{chong2006water,trout2019}.  When water is curtailed due to
shortages, it is made available to only those water rights holders who are
above a seniority threshold chosen by appropriate authorities
based on the drought severity. This naturally partitions the set of farmers
into two groups: potential sellers and potential buyers, leading to a
two-sided market. An example is provided in Figure~\ref{fig:example}.  In
addition, there are scenarios involving external entities, such as water
aggregators or brokers, who pool water from multiple sellers (see,
e.g.,~\cite{westernwatermarket,leonard2019expanding,selman2009water}) and
sell it to buyers.

Under this backdrop, we focus on a class of \emph{value-based}
resource matching problems. We consider a set of agents (farmers), each
associated with a discrete ordered set of resources (unit volumes of water,
or simply \emph{water units}). Each water unit is associated with a
\emph{value}, which depends on what use that unit of water is being put to
(low-value or high-value crops). Since farmers can have multi-crop portfolios,
the value of water units may vary, not only from one agent to
another, but also within an agent's resource set. Broadly,
our work is applicable to many market settings that involve multiple
identical units of resources, such as financial markets, electricity,
CPU job scheduling, and
bandwidth allocation~\cite{sandholm2002optimal,deng2007walrasian,khorasany2020paving}.

In a two-sided market where the agents are partitioned into sellers and
buyers by seniority, a seller's value for a water unit can be considered
the minimum price the seller is willing to accept, while, for a buyer, it
is the maximum price the buyer is willing to pay. Additionally, due to
geographic and legal constraints, not every buyer is \emph{compatible} with
a seller for trading water. This relationship is represented by a
buyer--seller bipartite compatibility graph; see Figure~\ref{fig:example}
for an example of stream flow and the resulting compatibility graph. The
objective is to obtain a \emph{trading assignment}, that is, a matching of
sellers' resources with buyers' needs, subject to compatibility and value
constraints.  We will assume that the agents are truthful about their valuations.  Every trading
assignment is assessed by the \emph{total social welfare} it generates~\cite{liu2016optimizing,xu2018two}.

\subsection{Contributions}
\para{Maximizing welfare under monotonicity constraints} We consider the
resource matching problem (\maxwelfare), where the goal is to maximize the
total welfare. We show that if the value functions satisfy certain
monotonicity properties~(under the assumption that every agent is rational
or
profit-maximizing~\cite{raffensperger2017,burke2004water,dang2016theoretical}),
an optimal matching can be obtained in polynomial time. We achieve this by
transforming the trading assignment problem into the maximum weighted
matching problem on a bipartite graph. To complement the above result, we
show that \maxwelfare{} is \cnp-hard even when the monotonicity constraints
are violated only on the buyer's side.


\para{Maximizing welfare under fairness constraints} 
We consider the problem
(\maxwelfarefair{}) of maximizing
welfare with the additional constraint that, for specified subsets of
buyers (where the subsets may also be singletons), a minimum number of water units must be assigned.
Such constraints can be viewed as a form of enforcing demographic fairness.
In general, we show that the problem of determining
whether there is an assignment that satisfies all 
the lower bound constraints is itself \cnp-complete.
When a solution satisfying all the constraints is known to exist and the value functions satisfy
monotonicity properties,  we present an efficient randomized algorithm
to find a solution that maximizes welfare  
and satisfies the given lower bound constraints
in expectation. 
To obtain this result, we use the dependent rounding algorithm of~\cite{gandhi2006dependent} and
leverage monotonicity properties of value functions.

\para{Maximizing Leximin satisfaction} We consider
the special case of a single seller--multiple buyers where each buyer
specifies the required number of water units. The \emph{satisfaction level} of a
buyer is the fraction of her requirement that is allocated. We consider the
problem where the objective is to find a trading assignment that maximizes
the satisfaction level of the least satisfied buyer. We provide an
efficient algorithm that finds a valid trading assignment satisfying the
following desirable properties: (i)~it maximizes the number of resources
matched over all valid assignments (thus maximizing seller profit), and
(ii)~in leximin order~\cite{Moulin-2004}, the vector of buyer satisfaction
levels is at least as large as that of any other valid assignment. The use of leximin order as a fairness criterion has
also been studied in several other contexts (see,
e.g.,~\cite{Chen-Liu-2020,Narang-etal-2022}).


\para{Experiments} We present results from experiments with a class of synthetic
data sets and two real-world data sets. The latter correspond to two water
basins in the state of Washington, US.  We study the impact of
factors, such as drought severity, value functions, and farmer seniority, on
the compatibility graph structure and objectives 
of trading assignments 
(e.g., welfare
maximization, maximizing satisfaction levels of buyers). 
Our results show that the combined effects of seniority, crop
profile, and geographic constraints can lead to varied trade outcomes
across different datasets.

\section{Related work}
\label{sec:related_work}
Many works have proposed mechanisms for optimal matching
of buyers to sellers in the context of water markets. 
Xu~et~al.~\cite{xu2018two} consider a two-sided market with simpler linear
value functions where they first apply weighted bipartite matching to
achieve welfare maximization. Then, they set transaction prices for each
assignment in the matching. In a single-seller/multiple-buyers framework,
Raffensperger and
Milke~\cite{raffensperger2009deterministic,raffensperger2017} use a
multipart bidding framework where a buyer's willingness to pay is modeled
as a monotone non-increasing function of the volume of water traded. They
develop a linear programming formulation that maximizes the consumer
surplus. Using a similar framework but accounting for water quality,
Sharghi and Kerachian~\cite{sharghi2022uncertainty} propose a multi-agent
optimization model. Noori~et~al.~\cite{noori2021agent} incorporate fairness
criteria into their models by requiring that each buyer should receive a
certain minimum amount of water depending on the  buyer's demand. The above
papers typically also account for a variety of agricultural and
socio-economic factors, resulting in very complex optimization problems
that are solved using heuristics.  As mentioned earlier, our work is
applicable to settings that involve indistinguishable units of a resource.
In such a setting, Sandholm and Suri~\cite{sandholm2002optimal} consider
the problem of optimal clearing where sellers and buyers specify bids
through supply and demand curves.

To our knowledge, very few papers have addressed computational aspects of
water markets. Liu~et~al.~\cite{liu2016optimizing} consider the problem of
optimal trading assignments in water right markets. They consider two
maximization objectives--social welfare  and flow--with a minimum threshold
constraint on the volume of water traded in each transaction. They consider
a setting with linear value functions, where the problem of maximizing
welfare can be viewed as maximizing flows in a weighted seller--buyer
bipartite graph.  Li~et~al.~\cite{li2017stability} consider the same
setting and examine the assignment problem from the perspective of
cooperative game theory.  Both works present experimental results using
water market data.

Our work is also related to resource allocation problems that are modeled
as multi-round matchings~\cite{trabelsi2023resource} and repeated
matchings~\cite{caragiannis2023repeatedly}.  
In both cases, the problem can be viewed as a
matching (or a $b$-matching) problem~\cite{LP-1986} on a bipartite graph,
with multiple copies of nodes corresponding to each resource or agent.
This is similar to our work, where the water units corresponding to each
seller or buyer are represented as nodes of a bipartite graph.

Several recent papers have addressed fairness issues in bipartite matching. For example,
Lesmana~et~al.~\cite{lesmana2019balancing} develop an algorithm with
provable guarantees for the trade-off between operator benefit (in our
case, a single seller) and the minimum satisfaction or utility for the
customer (in our case, a buyer).
Esmaeili~et~al.~\cite{esmaeili2023rawlsian} consider Rawlsian fairness in
online bipartite matchings.  Methods to achieve group fairness have also
been considered in both offline and online versions of the bipartite
matching problem~\cite{ma2022group,esmaeili2023rawlsian,panda2022bipartite}.


\section{Preliminaries}
\label{sec:preliminaries}
Let~$[k]$ denote the set $\{1,2,\ldots,k\}$ and~$\nonnegreals$ be the set
of nonnegative real numbers.

\para{Agents, resources, and value functions}
Let~$A=\{a_1,a_2,\ldots,a_N\}$ denote the set of $N$  agents. The agents are
ordered by seniority;~$a_i$ is senior to~$a_j$ for any~$i>j$.
Each
agent~$a\in A$ is associated with an ordered set
$\Gamma_a=\{w^a_1,w^a_2,\cdots,w^a_{\gamma_a}\}$ of ~$\gamma_a$ resources
or water units.  The elements of  $\Gamma_a$ are ordered so that for all $x
< y$, water unit $w^a_x$ must be sold or bought before the water unit
$w^a_y$; we use the notation $w^a_x \prec{} w^a_y$ to indicate this
ordering.
A value function~$\valfun_a:\Gamma_a\rightarrow\nonnegreals$
assigns a nonnegative value~$\valfun_a(w)$ to each~$w\in\Gamma_a$.  Each water
unit of an agent can be considered to be associated with a specific use
(e.g., crop type, which field it is applied to in a farm), and its value
can correspond to the anticipated profit, its importance to keep the crop alive or
healthy, etc. (see, e.g.,~\cite{raffensperger2017}). The example in
Figure~\ref{fig:example} shows agents with associated resources and value
functions.

\para{Buyers, sellers, and trading} Depending on water availability, the
agent set~$A$ is partitioned into two sets, namely sellers~$S$ and
buyers~$B$.  We let $N_S = |S|$ and $N_B = |B|$.  Each seller~$s$
has~$\gamma_s$ water units, which is the agent's capacity, while each
buyer~$b$ has a requirement of~$\gamma_b$ water units.  A \emph{trading
assignment}~$\trade$ consists of a matching of buyer water units with
seller water units; it is specified by a set of ordered pairs of the
form~$(w^s_i,w^b_j)$.

\para{Compatibility} A seller~$s$ is \emph{compatible} with a buyer~$b$ if
$b$ is allowed to use the water right owned by~$s$.
This compatibility relationship is determined by
geographic factors such as whether they share a common stream and
prevailing water laws.  This relationship is represented by a seller--buyer (undirected)
bipartite compatibility graph~$G(S,B,E)$; a seller $s \in S$ is compatible
with a buyer $b \in B$ if and only if there is an edge between $s$ and $b$
in~$G$. The example in 
Figure~\ref{fig:example} (third panel)
shows a compatibility graph induced by the geographic positions of agents and water
availability. 

\para{Total value and welfare from trade} We assume that every water unit
will be used regardless of whether it is traded or not. If a water unit is
traded, then it is used by the corresponding buyer; otherwise, it is used
by the seller. The value extracted from each water unit will depend on who
uses it (the seller or buyer) and how it is used. For example, if a
seller's unit~$w^s_i$ is matched to a buyer's unit~$w^b_j$, then its new
value is~$f_b(w^b_j)$. The \emph{total value} before trade
is~$\totvalnotrade=\sum_{s\in S}\sum_{w\in \Gamma_s}f_s(w)$. Given a
trading assignment~$\trade$, let~$W_\trade$ denote the set of matched
resources of sellers and let $\overline{W_\trade}$ denote the set of
unmatched resources of sellers. The total value for a given trading
assignment~$\trade$ is $\totval(\trade) = \totvalnotrade{} +
\welfare(\trade)$ where $\welfare(\trade)= \sum_{(w^s_i,w^b_j)\in
\trade}[f_b(w^b_j)-f_s(w^s_i)].$
Note that the welfare function does not account for profits of individual
agents, which is determined by the transaction price for each trade.

\begin{remark}
\label{rem:assumptions}
Here, we assume that the value functions are public. 
We also assume that the value of each water
unit remains the same regardless of the role (buyer or seller) of the agent
associated with it. In general, this need not be the case. For
example, if an agent risks losing a crop that corresponds to a multi-year
investment, she might be willing to pay much more for the water than the
annual value of the crop.  We also assume that every agent participates in
the market as a seller or a buyer. This also need not be the case in real-life;
for example, some farmers are known to exhibit non-pecuniary
behavior~\cite{reetwika2023thesis,cook2014assessing}, 
that is, they would
reduce their gains by opting to farm rather than sell their water.
\end{remark}

\section{Maximizing welfare}
\label{sec:welfare}

\subsection{Problem Definition}
We now define a welfare maximizing resource matching problem where the goal
is to match sellers' resources to buyers' needs such that, for each agent,
the matching respects the value-based ordering of the resources, i.e., if a
resource is matched in a solution, then all units valued higher than this
resource in the agent's portfolio are also matched. Also, for every matched
resource, the value assigned to it by the seller is at most the value
assigned by the buyer.
\begin{problem}[Maximum Welfare Water Trading problem -- \maxwelfare{}] 
\label{prob:maxwelfare}
Given sets of sellers~$S$, buyers~$B$, their water units, associated value
functions, and a compatibility graph~$G(S,B,E)$, find a trading
assignment~$\trade^*$ that maximizes the welfare function~$\welfare(\cdot)$
subject to the following constraints: 
(i)~Buyer values the unit at least
as much as the seller: for every matched pair~$(w^s_i,w^b_j)$ where~$w^s_i$
is the $i$th unit of seller~$s$ and~$w^b_j$ is the $j$th unit of buyer~$b$,
$\valfun_b(w^b_j)\ge\valfun_s(w^s_i)$, and (ii)~Matching is consistent
with ordering of resources: for any agent~$a\in S\cup B$ and~$i>1$,~$w^a_i$
is matched only if~$w^a_{i-1}$ is matched.
\end{problem}
\noindent
Henceforth, we refer to a trading assignment that satisfies the two
conditions above as a \emph{valid trading assignment}.

\subsection{Monotone Value Functions}
\label{sec:monotone}
Here, we show that, with certain monotonicity 
constraints on  value
functions, the \maxwelfare{} problem can be solved efficiently. 
The conditions are as follows.
For each seller $s$, the value function is 
monotone non-decreasing (i.e., 
$\valfun_s(w^s_i) \geq \valfun_s(w^s_{i-1})$ for all $i > 1$) and,
for each buyer $b$, the value function is monotone non-increasing 
(i.e., 
$\valfun_b(w^s_i) \leq \valfun_b(w^s_{i-1})$ for all $i > 1$).
These correspond to rational or
profit-maximizing agents; a
seller would sell the first assigned resource that was
meant for the lowest valued use while
a buyer will use the first assigned resource
for the highest valued use.
\begin{theorem}\label{thm:monotone}
Suppose we are given a set of sellers~$S$, buyers~$B$,
their respective water units, a
compatibility
graph~$G(S,B,E)$, and value functions satisfying the following
criteria: 
$\forall s\in S$, $\valfun_s$ is a monotone non-decreasing
function and
$\forall b\in B$, $\valfun_b$ is a monotone non-increasing
function. 
In this setting, \maxwelfare{} can be solved in time
polynomial in the total number of water units.
\end{theorem}
\noindent
We show that Algorithm~\ref{alg:monotone} solves \maxwelfare{} for monotone
value functions. We start with the following definition.

\para{Resources--needs compatibility graph} Given the
compatibility graph~$G(S,B,E)$ and the value functions, we construct
an edge-weighted bipartite graph~$G'(S', B', E')$ as follows. 
For each seller
water unit~$w^s_i$ of agent~$s\in S$, we create a node~$s_i$ in $G'$.
For each buyer water
unit~$w^b_j$ of agent~$b \in B$, 
we create a node~$b_j$ in $G'$.
Let~$B'=\{b_j\mid \forall b\in B,~w^b_j\in\Gamma_b\}$ and~$S'=\{s_i\mid
\forall s\in S,~w^s_i\in\Gamma_s\}$. The edge set $E'$ is defined as
follows:~$(s_i,b_j)\in E'$ if and only if (i)~$b$ is compatible with~$s$ in~$G$ (i.e.,~$\{s,b\}\in E$) and
(ii)~$\valfun_b(w^b_j)\ge\valfun_s(w^s_i)$. 
The weight $\alpha_e$ on each
edge~$e=\{s_i,b_j\}$ is given by 
$\alpha_e=\valfun_b(w^b_j)-\valfun_s(w^s_i)$.
(See the rightmost panel in Figure~\ref{fig:example}
for an example.)

\newcommand{\cust}{\vspace{0em}}
\SetAlgoSkip{cust}
\begin{algorithm}[ht]
\caption{\maxwelfare{} with monotone value functions.}
\label{alg:monotone}
\DontPrintSemicolon
\BlankLine
\SetKwInOut{Input}{Input}
\SetKwInOut{Output}{Output}
\Input{Buyer set~$B$, seller set~$S$, compatibility graph~$G(S,B,E)$,
value functions~$\valfun_a(\cdot)$, $\forall a\in S\uplus B$ that satisfy
the conditions of Theorem~\ref{thm:monotone}.}
Construct resources--needs compatibility graph~$G'(B',S',E')$.\;
Find a maximum weighted matching~$\match$ of~$G'$.\;
Let trading assignment~$\trade=\varnothing$.\;
\For{$(s_i,b_j)\in \match$}{
    Add $(w^s_i,w^b_j)$ to $\trade$.\;
}
\While{$\exists a\in B\uplus S$ and $\exists i>1$~s.t. $w^a_i$ is matched
in $\trade$ but $w^a_{i-1}$ is not}{
    \label{line:while}
	Replace~$w^a_i$ with~$w^a_{i-1}$ in $\trade$.
}
\Output{Trading assignment~$\trade$}
\end{algorithm}
\begin{proof}[Proof of Theorem~\ref{thm:monotone}]
For a matching~$\match$ in $G'$, let $\weight(\match)$ denote the sum of
the weights of all the edges in~$\match$. Note that any trading assignment
$\trade$ corresponds to a unique matching in~$G'$: $(s_i,b_j)\in \match$ if
and only if~$(w^s_i,w^b_j) \in \trade$. Also,~$\weight(\match)= \sum_{e
\in \match}\alpha_e = \sum_{e=\{s_i,b_j\}} [\valfun_b(i)-\valfun_s(j)] =
\welfare(\trade)$. Hence, the maximum welfare that can be achieved from
any~$\trade$ is at most the weight of any maximum weighted
matching of~$G'$.

We now show that the output of Algorithm~\ref{alg:monotone}
satisfies the priority constraints defined in
Problem~\ref{prob:maxwelfare}. Since~$\match$ is a maximum weighted
matching,~$\trade$ corresponds to an assignment with maximum welfare.
However, it might not satisfy the priority constraints stated in
Problem~\ref{prob:maxwelfare}.  Suppose~$\trade$ is not a valid trading
assignment, i.e., $\exists a\in B\uplus S$, such that, for
some~$i>1$,~$w^a_i$ is matched but~$w^a_{i-1}$ is not.  Without loss of
generality, let~$a\in B$.  Let~$(w^s_i,w^a_j)\in \trade$.
Let $\trade'$ = $(\trade \setminus \{(w^s_i,w^a_j)\}) \cup
\{(w^a_{i-1},w^b_j)\}$.  Since $(w^s_i,w^a_j)\in \trade$,~$a$ is compatible
with~$s$, i.e., $(s,a)\in E$ and~$\valfun_a(w^a_j)\ge\valfun_s(w^s_i)$ by our
construction of the bipartite graph~$G'$. Since ~$\valfun_a$ is a monotone
non-increasing function,~$\valfun_a(w^a_{j-1})\ge\valfun_a(w^a_j)$. Hence, $\trade'$ is also a valid assignment.
Also, since~$\valfun_a(w^a_{j-1})-\valfun_s(w^s_i)\ge\valfun_a(w^a_j)-\valfun_s(w^s_i)$, $\welfare(\trade')\ge\welfare(\trade)$, which implies that the new
assignment does not reduce welfare. (In fact, the welfare cannot increase
since $\trade$ has the maximum welfare value.) The same argument holds for
every iteration in the loop defined on Line~\ref{line:while} in the
algorithm. Noting that the maximum weighted bipartite matching can be
computed in polynomial time, the theorem follows. For details on the
running time, see supplement.
\end{proof}

\para{Non-monotone value functions}
One may ask whether an efficient algorithm is possible under weaker
assumptions on the value functions.  We now consider a version of the
\maxwelfare{} problem where the value functions for sellers are monotone
non-decreasing, while those for the buyers are \underline{not} required to
satisfy the monotone non-increasing property.  Our next result points out
the complexity of the \maxwelfare{} problem for that setting.
\begin{theorem}\label{thm:hard_dis_buyer_nonmonotone}
Given a set of sellers~$S$, buyers~$B$, their water units,
compatibility graph~$G(S,B,E)$, and value functions satisfying the
following condition: $\forall s\in S$, value function $\valfun_s$ is a
monotone non-decreasing function, \maxwelfare{} is
\cnp-hard.
\end{theorem}
\noindent
\textbf{Proof Idea:}
Our reduction is from the 
\textsc{Exact Cover by 3-Sets} problem;
see Section~\ref{app_sse:hardness_buyer} of
the supplement.

\begin{remark}
\label{rem:threshold}
By examining our proof of Theorem~\ref{thm:hard_dis_buyer_nonmonotone}, it
can be seen that the \maxwelfare{} problem is hard when value functions
(for both the sellers and buyers) are threshold functions (i.e., they have
a non-zero value only when the number of water units sold by a seller or
assigned to a buyer is at least a given positive integer).  Thus, the
problem of maximizing welfare is \cnp-hard when agents have a lower bound
on the number of water units they sell/buy before a trade provides value to
an agent.
\end{remark}

\section{Resource matching with fairness}
\label{sec:fairness}
Here, we present two results incorporating fairness criteria corresponding
to the buyers.  The first result is on maximizing welfare subject
to lower bounds on the number of water units assigned to groups of buyers.
The second problem addresses Leximin fairness, which is a generalization of
Rawlsian fairness.

\subsection{Buyers' Lower Bound Fairness Constraints}
\label{sse:maxwelfarefair}

Let~$\subsetcol=\{L\mid L\subseteq B\}$ be a collection of
subsets\footnote{In general, $|\subsetcol|$ can be exponential in $|B|$. We
will assume that $|\subsetcol|$ is bounded by a polynomial in $|B|$.} of
buyers. For each~$L \in \subsetcol{}$, let~$r(L)$ be a positive integer
denoting the minimum (total) number of water units to be assigned to the
buyers in~$L$. Note that~$L$ can correspond to a single buyer as well.


\begin{problem}[Maximum Welfare Water Trading with buyers' lower bound Fairness constraints
(\maxwelfarefair)] 
\label{prob:maxwelfare-fair}
Given sets of sellers~$S$, buyers~$B$, associated value functions,
collection of subsets~$\subsetcol$, function~$r:\subsetcol\rightarrow
\posints$, and a compatibility graph~$G(S,B,E)$, find a trading
assignment~$\trade^*$ that maximizes the welfare function~$\welfare(\cdot)$
under the following constraints: (i)~for every matched
pair~$(w^s_i,w^b_j)$ where~$w^s_i$ is the $i$th unit of seller~$s$
and~$w^b_j$ is the $j$th unit of buyer~$b$,
$\valfun_b(w^b_j)\ge\valfun_s(w^s_i)$, (ii)~for any agent~$a\in S\cup B$,
and~$i>1$,~$w^a_i$ is matched only if~$w^a_{i-1}$ is matched, and (iii) for
each $L\in\subsetcol$, the number of water units assigned to~$L$ is at least 
$r(L)$.
\end{problem}


It is easy to construct
instances where there is no solution that satisfies all the lower bound
constraints. This also implies that buyers' lower bound fairness constraints  can
arbitrarily affect the welfare objective.  When the subsets for which lower bound constraints specified
are pairwise disjoint, the feasibility problem shares some similarity with
the construction of coalitions to optimize certain functions of agents'
utilities in hedonic games (see
e.g.,~\cite{Waxman-etal-2021,Waxman-etal-2020}) and the problem of
partitioning the node sets of graphs so that the subgraph induced on each
block of the partition has a specified  minimum degree (see,
e.g.,~\cite{Alon-2006,Bang-Jensen-etal-2016,Stiebitz-1996}).  It should be
noted that \maxwelfarefair{} also involves matching-related constraints.
For this problem, \emph{whenever a solution which satisfies all the
constraints exists}, we show below  (Theorem~\ref{theorem:maxwelfare-fair})
that there is a polynomial time randomized algorithm to maximize the
welfare.
\begin{theorem}
\label{theorem:maxwelfare-fair}
Let $\mathcal{M}$ denote the set of all trading assignments $\mathcal{T}$
which satisfy  the lower bound constraints associated with all $L\in
\subsetcol$. Suppose $\mathcal{M}\neq\emptyset$.  If $f_s$ and $f_b$ have
the same monotonicity properties as in Theorem~\ref{thm:monotone}, there is a polynomial time randomized
algorithm to find a trading assignment $\mathcal{T}$ 
satisfying the following properties: (1)~(Single buyer constraint) for each~$L\in
\subsetcol$ such that~$|L|=1$, the amount of water assigned is at
least~$r(L)$; ~(2)~(Demographic constraint) for~each $L \in \subsetcol{}$ where $|L|>1$, the lower bound
constraint for $L$ is satisfied in expectation, and (3) the expected welfare of\/
$\mathcal{T}$ is at least the maximum welfare among all the assignments in
$\mathcal{M}$.
\end{theorem}
\begin{proof}
Let~$G'(S',B',E')$ be the resources-needs compatibility graph.  Recall that
in~$G'$,~$s_i$ represents the~$i$th water unit of seller~$s$, and~$b_j$
represents the~$j$th water unit of buyer~$b$.  For a node~$v$
in~$G'$, let $N(v)$ denote the set of neighbors of~$v$.
Let~$\alpha_{sibj}=\valfun_b(w^b_j)-\valfun_s(w^s_i)$ be the weight on the
edge~$(s_i,b_j) \in E'$.  We formulate the following linear program (LP)
with a variable $z_{sibj}$ for each possible assignment $(s_i, b_j)$ in the
resource--needs compatibility graph.
\begin{align*}
\max \quad \sum \alpha_{sibj} z_{sibj}
\end{align*}
\begin{align}
\forall (s_i,b_j)\in E', \quad z_{sibj} \geq 0, \label{eqn:mat1}
\end{align}
\begin{align}
\forall b_j\in B',~\sum_{s_i\in N(b_j)}z_{sibj} \le 1;~\forall
s_i\in S',~\sum_{b_j\in N(s_i)}z_{sibj} \le 1,  \label{eqn:mat2}  
\end{align}
\begin{align}
\forall L\in\subsetcol,\quad 
\sum_{b\in L}\sum_{s, i, j}z_{sibj} \geq r(L)\,.
\label{eqn:fair}
\end{align}
The constraints in~(\ref{eqn:mat2})  
correspond to matching constraints, while those 
in~(\ref{eqn:fair}) capture the fairness conditions.
Because of our assumption regarding feasibility, there is an optimal
fractional solution $z$ to the above LP.  Note that the solution $z$ will
satisfy the property that if $z_{bisj}>0$, it must be the case that
$\sum_{s'j'} z_{b(i-1)s'j'}=1$. Otherwise, we can modify the fractional
solution in the same way as in the proof of Theorem~\ref{thm:monotone} and
achieve this property.

We use the dependent rounding algorithm of~\cite{gandhi2006dependent},
which rounds each $z_{bisj}$ to an integer variable $Z_{bisj}$ such that
$\Pr[Z_{bisj} = 1] = z_{bisj}$, and $\sum_{i,s,j} Z_{bisj} \in \{\lfloor
\sum_{i, s, j} z_{bisj} \rfloor, \lceil \sum_{i, s, j} z_{bisj} \rceil \}$.
Since~$r(\cdot)$ is an integer, for~$L=\{b\}$, it follows that $\lfloor
\sum_{i, s, j} z_{bisj} \rfloor\geq r(\{b\})$, so that $\sum_{i,s,j}
Z_{bisj}\geq r(\{b\})$.  Additionally, $\mathbb{E}[\sum_{b\in L} \sum_{i,
s, j} z_{bisj}] \geq r(L)$ for each~$|L|>1$. In a similar manner,
the expected value of the objective function is at least the objective
value of the LP.
\end{proof} 

Note that the solution $\mathcal{T}$ from
Theorem~\ref{theorem:maxwelfare-fair} need not satisfy the lower bound
constraints for a given~$L\in \subsetcol$ -- it is only satisfied in
expectation, over the random choices made by the algorithm.  It actually gives a (fractional)
solution whenever the LP is feasible, which might happen even if
$\mathcal{M}=\emptyset$. We note below that if the \maxwelfarefair{} instance
has lower bound constraints only for individual buyers,
they are satisfied exactly as the resulting
LP represents an instance of the  
$b$-matching\footnote{See Section~\ref{app_sec:intro} of the supplement for a definition of the $b$-matching problem.}
problem.

\begin{corollary}
\label{corollary:maxwelfare-fair}
Let $\mathcal{M}$ denote the set of all trading assignments $\mathcal{T}$
which satisfy the lower bound constraints associated with all $b\in B$.
Suppose $\mathcal{M}\neq\emptyset$.  If $f_s$ and $f_b$ have the same
monotonicity properties as in Theorem~\ref{thm:monotone}, it is possible to
find a trading assignment $\mathcal{T}$ in polynomial time, which ensures
that: (1)~for each buyer $b\in B$, the amount of water assigned is at least
$r(\{b\})$, and (2) the expected welfare of\/ $\mathcal{T}$ is at least the
maximum welfare among all the assignments in $\mathcal{M}$.
\end{corollary}

\para{Complexity} We show that, in general, the problem of determining
whether there is a matching solution that satisfies all the lower
bound constraints is itself an NP-complete problem. 
This is shown using a reduction from
the \textsc{Minimum Vertex Cover} problem 
(see Proposition~\ref{pro:feasdemog_hard} in Section~\ref{app_sse:maxwelfarefir} of the
supplement).



\subsection{Leximin Fairness}
\label{sse:leximin_fair}

We consider a simpler setting of a single seller with a set of~$k$
resources or water units $W=\{w_1,w_2,\ldots,w_k\}$ and multiple buyers
with requirements.  All water units have the same value. 
The seller's
objective is to maximize the number of resources sold, 
subject to the constraints
represented by a resource--buyer compatibility 
graph~$G_w(W,B,E_w)$ and
an additional fairness condition discussed below.
For any assignment~$\trade$ and buyer~$b\in B$, let~$\eta_b$ denote the total number
of water units assigned to~$b$.  Let~$\gamma_b$ be the number of units required by~$b$.  
Let~$\psi(\trade)=(\eta_b/\gamma_b\mid b\in B)$ denote
the \emph{buyer satisfaction vector} corresponding to~$\trade$. 
The fairness condition we impose is based on \textbf{leximin ordering} of vectors defined 
below.

\noindent
\textbf{Leximin Ordering:}~
Suppose we
have two real sequences,  
$\mu_1$ and $\mu_2$, each of length~$k$.
We say that $\mu_1$ is \emph{leximin
larger} than $\mu_2$ if there exists an integer~$0\leq\ell<k$ such that the
first~$\ell$ smallest elements of both vectors are equal, while the
$(\ell+1)$-smallest element of $\mu_1$ is greater than the
$(\ell+1)$-smallest element of $\mu_2$. 

Suppose the sequences $\mu_1$ and $\mu_2$ represent the satisfaction vectors of buyers created
by two assignments $\trade_1$ and $\trade_2$ and $\mu_1$ is
leximin larger than $\mu_2$.
From a fairness perspective, $\trade_1$ 
is preferable since, for some integer $\ell{} < k$, the 
$(\ell+1)$th least satisfied buyer has a larger satisfaction 
level in $\mu_1$ compared to that in $\mu_2$. 
(For all lower values, the satisfaction ratio of buyers in $\mu_1$
is at least as large as that of $\mu_2$.)
This motivates the following problem.

\noindent
\begin{problem}[\maxleximin{}] 
\label{prob:leximin}
Given a seller with a set $W$ of $k$ water units, a set of buyers~$B$ with the same cost for every
water unit, and a compatibility 
graph~$G_w(W,B,E_w)$, find a trading
assignment~$\trade^*$ with the leximin-largest buyer satisfaction vector.
\end{problem}
\begin{theorem}
\label{thm:leximin}
An optimal solution to the \maxleximin{} problem can be obtained in polynomial time.
\end{theorem}
\noindent 
\textbf{Proof outline:}
First, we show that an instance of
\maxleximin{} can be reduced to an instance of the multi-round matching
problem called~\maxtbmrm{} from Trabelsi~et~al.~\cite{trabelsi2023resource}. In
multi-round matching,~$X$ is a set of agents and~$Y$ is a set of resources
where agents need to be matched to resources in~$k$ rounds for some positive
integer~$k$. A bipartite compatibility graph~$G_m(X,Y,E_m)$ indicates which
resource is compatible with which agent for matching. Each agent~$x_i$ has a
permissible set of rounds~$K_i\subseteq[k]$ in which it can be matched,
and~$\rho_i\le |K_i|$ is the desired number of rounds in which it will be
matched. In addition,~$\mu_i:[\rho_i]\rightarrow \posreals$ is a benefit
function for agent~$x_i$, which gives a benefit value~$\mu_i(\ell)$ when
the number of rounds assigned to~$x_i$ is~$\ell$. The objective is to
find a $k$-round matching to maximize the total benefit. Given an instance
of \maxleximin{}, we can construct an instance of \maxtbmrm{} as follows.
\begin{enumerate}[leftmargin=*,noitemsep,topsep=0pt]
\item Construct a new compatibility graph~$G_m(X,Y,E_m)$ where~$X=\{v\}$ is
a special node that represents a specific resource in every round,
$Y=B$, the set of buyers, and~$E_m=\{(v,b)\mid b\in B\}$.
($G_m$ is a star
graph with center~$v$ and nodes of~$B$ as leaves.)
\item The number of rounds~$k=|W|$, one for each water unit in~$W$.
\item For each buyer~$b$, $K_b=\{i\mid (w_i,b)\in E_w\}$, i.e., the rounds
correspond to those representing compatible resources,
and~$\rho_b=\gamma_b$, the requirement of~$b$. The construction of the
benefit function~$\mu_b$ follows the construction used in Theorem~4.9 in~\cite{trabelsi2023resource}.
\end{enumerate}
\noindent
Given a multi-round matching solution~$\match$ to the above instance of
the~\maxtbmrm{} problem, we construct a trading assignment~$\trade$ as
follows: Each matching edge~$(v,b)$ corresponds to some round~$i$. It is 
mapped to the assignment~$(w_i,b)\in\trade$. The proof that the solution
satisfies leximin-largest criterion uses several additional results; it is presented in 
Section~\ref{app_sec:leximin_fair} of the supplement. 

\begin{remark}
We note that in Trabelsi~et~al.~\cite{trabelsi2023resource}, it was only
shown that their solution for the relevant benefit function satisfies the
Rawlsian social welfare, i.e., the solution maximizes the satisfaction of
the least satisfied buyer. Here, in the context of trading
assignments, we show that the same construction
provides a stronger leximin-largest solution.
\end{remark}

\section{Experiments}
\label{sec:exp}

We experimented with real-world and synthetic datasets\footnote{ The data
and the code for running the experiments are available in
Github~\cite{datacode}. The data is summarized in Table~\ref{tab:data} in
Section~\ref{app_sec:rw-dataset} of the supplement.} to study resource
matching under various scenarios determined by drought severity, types of
value functions, and agent seniority. All our experimental results rely on
the assumptions mentioned in Remark~\ref{rem:assumptions}.


\begin{figure*}[tb]
\centering
\includegraphics[width=\textwidth]{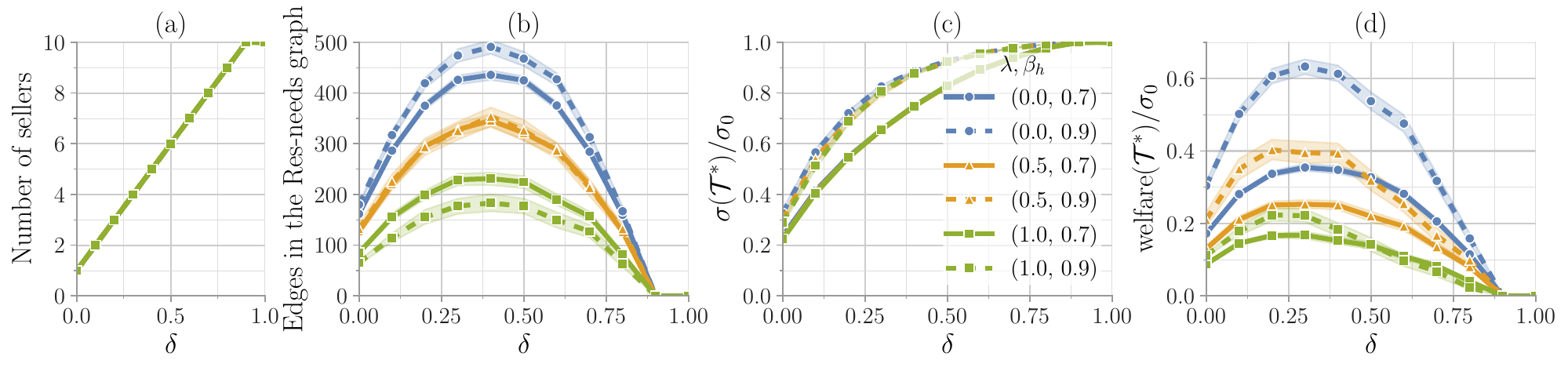}
\caption{\normalfont Panels (a) and (b) show structural properties of the
    resources--needs graph for the synthetic datasets with respect to increasing water availability. 
In Panels (c) and (d), Y-axis gives the total value~$\totval(\trade^*)$ and~$\welfare(\trade^*)$ corresponding to an optimal assignment~$\trade^*$ respectively, normalized by~$\totval_0$, the total value when 100\% of water is available. 
All
results are for~$N=10$ and the number of units per agent~$k=5$. The results are
shown for different values of~$\lambda$ 
and~$\betahigh$.}
\label{fig:synthetic}
\end{figure*}

\subsection{Datasets}
\para{Synthetic datasets}
We consider a simple setup where there are~$N$ agents~$A=\{a_i\mid 1\le
i\le N\}$ where a buyer can buy from any seller as long as there is value
compatibility.  From a domain perspective, this setup models the situation
where there is a single stream, and, therefore, an agent can potentially
access any other agent's water.  We will assume that all capacities and
requirements are the same; that is, for all agents~$a\in S \cup
B$,~$\gamma_a=k$. Water availability~$\delta$ determines the fraction of
water that is available.  If~$\delta=1$ (similarly,~$\delta=0$), then all
(none of the)~$N k$ units of water are available, and therefore, there is
no trade. Every agent $a_i$ is associated with~$k$ units of water.
Given~$\delta$, agent~$a_i$ is a seller if and only
if~$\frac{i}{N}\ge1-\delta$ (larger the~$i$, the higher the priority).  Agents are categorized into two types:
high-valued and low-valued. For a
seller~$s$, we will consider a simple linear value
function:~$\valfun_s(\ell)=\beta\ell$ for~$\ell=1,\ldots,k$; 
high-valued agents have a larger~$\beta$ than low-valued agents.
Similarly, for a buyer~$b$, we will consider the following linear
function:~$\valfun_b(\ell)=\beta(k-\ell+1)$ for~$\ell=1,\ldots,k$. To
decide whether agent~$a_i$ is high-valued or low-valued, we will define a
probability function as follows: $p_h(a_i) = \lambda
\frac{i}{N}+(1-\lambda)\big(1-\frac{i}{N}\big)$, where~$\lambda\in[0,1]$ is
a tunable parameter. The higher the~$\lambda$, the greater the probability
that high priority agents are high-valued. If~$\lambda=0.5$, then all
agents have the same probability of~$0.5$ to be assigned to the high-valued
category. In our experiments, the~$\beta$ value for the high-valued agent,
denoted by~$\betahigh$, is a real value between~$0.5$ and~$1$, and
for ~$\betalow$, the~$\beta$ for the low-valued agent, is set
to~$1-\betahigh$.

\para{Real-world datasets}
We used datasets containing~93 usable water rights held in the Touchet
River Watershed and~77 usable water rights held in the Yakima River
Watershed in the state of Washington, along with their associated farm
attributes, including acreage, crop types, and volume of water needed. The
value of a water unit was calculated based on value of production per acre
for the relevant crop types~\cite{nass2022} ($n_i$) and the volume of water
required for each field ($v_i$).  Then, we calculated the value per
acre-foot $p_i$ = $A_i\cdot n_i$/$v_i$, where $A_i$ is acreage, $n_i$ is
the value of production per acre based on crop type.  Buyers and sellers
were decided based on water availability and water right seniority.  The
aggregated volumes per field $v_i$ were disaggregated into prioritized
units of water (with unit sizes being $5$, $10$, or $20$ acre-feet). For
buyers, units were prioritized in descending order of their value, while,
for the sellers, units were prioritized in ascending order of their value,
thus satisfying the monotonicity constraints of
Algorithm~\ref{alg:monotone}.  We created the resources--needs bipartite
graph using the value functions and geographic locations of the water
rights. This is described in more detail in the supplement.

\subsection{Results}
\begin{figure*}[tb]
\centering
\includegraphics[width=\textwidth]{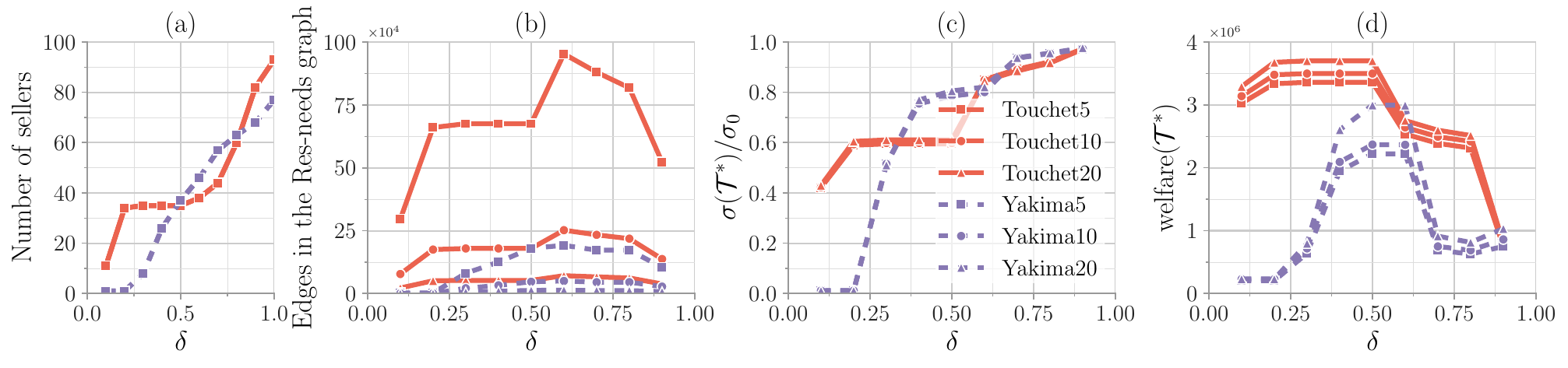}
\caption{\normalfont Panels (a) and (b) show structural properties of the resources--needs graph for real-world
datasets  with respect to increasing water availability. 
In Panels (c) and (d), Y-axis gives the total value~$\totval(\trade^*)$ and~$\welfare(\trade^*)$ corresponding to an optimal assignment~$\trade^*$ respectively, normalized by~$\totval_0$, the total value when 100\% of water is available.  All results are for varying sizes for water units.
}
\label{fig:real}
\end{figure*}
\para{Compatibility graph structure and water availability} 
Here, we examine the structure of the buyer--seller and resources--needs
compatibility graphs with increasing water availability~$\delta$.  For the
synthetic graphs, Figure~\ref{fig:synthetic}(a) shows a linear increase in
the number of sellers, which is due to the fact that each agent is assigned
the same number of resources. Note that the buyer--seller compatibility
graph in this case only differs with respect to~$\delta$, as all other
parameters only determine the value of the water units. However, the
resources--needs bipartite graph (defined in Section~\ref{sec:monotone}) is
influenced significantly by combinations of seniority and value. In
Plot~\ref{fig:synthetic}(b), we observe that the number of edges in the resources--needs graph
significantly decreases as the number of senior high-value agents ($\lambda$
and~$\betahigh$ being both high) increases due to the fact that most high-value agents have
water, while low-value agents who do not have water cannot buy from the
former group. The corresponding set of plots for real-world datasets are in
Figure~\ref{fig:real}. We note that, in this case, the number of sellers
in Plot~\ref{fig:real}(a) does not increase linearly with~$\delta$, particularly in
the case of the \touchet{} networks. The curve plateaus at
around~$\delta=0.2$ before rising at~$\delta>0.5$ in Plot~\ref{fig:real}(a).  The reason
for this is the existence of senior agents requiring very large numbers of
water units.  Until sufficient water is available, these agents will be
classified as buyers instead of sellers, causing the aforementioned
plateaus. Therefore, as~$\delta$ is increased, the number of available
water rights for trade increases abruptly. This also leads to the
plateauing in Plot~\ref{fig:real}(b).  Overall, we observe that heterogeneity (both
quantity and crop value) in crop portfolios, seniority, and geographic
constraints can lead to fewer compatible seller-buyer unit pairs. Also, the number of such
pairs is relatively low in the case
of \yakima. 

\para{Welfare from trade and water availability}
Figures~\ref{fig:synthetic}(c) and~(d) show the benefit of trading for synthetic datasets. The
total value due to trading is significantly higher when water availability
is around 50\%. We observe that the combination of seniority and crop value
(high or low) has a significant effect. A scenario corresponding to
high-value buyers and low-value sellers ($\lambda=0$) offers more
opportunities for matching than the other way round ($\lambda=1$). The
welfare peaks when~$\delta$ is in the interval~$[0.25,0.35]$, which is also the
interval with the highest number of edges in the bipartite graph. The
parameter~$\betahigh$ contributes significantly to the value of welfare.
The larger the~$\betahigh$, the greater the total welfare~$\totval(\trade)$.

However, in the case of real-world datasets, we see a much richer behavior,
which is partly explained by the structure of the resources--needs
network. The \yakima{} dataset exhibits characteristics similar to those of
synthetic datasets around~$\delta=0.5$, but the normalized total value
drops close to zero even when around~25\% of the water is available
(see Figure~\ref{fig:real}(c)). This is due to the same reason as 
that for the \touchet{} dataset: an agent
with a large number of water units. Only for a sufficiently large value
of~$\delta$ does this agent get to exercise its water unit and become a
seller. For the \touchet{} data, we observe the same phenomenon as was
observed for the number of edges at~$0.25\le \delta\le 0.5$.
We note that the welfare in the case of the \touchet{} networks is
much larger than that for \yakima{}, where
seller-buyer compatibility is relatively low.
\begin{figure}[htb]
    \centering
    \includegraphics[width=\columnwidth]{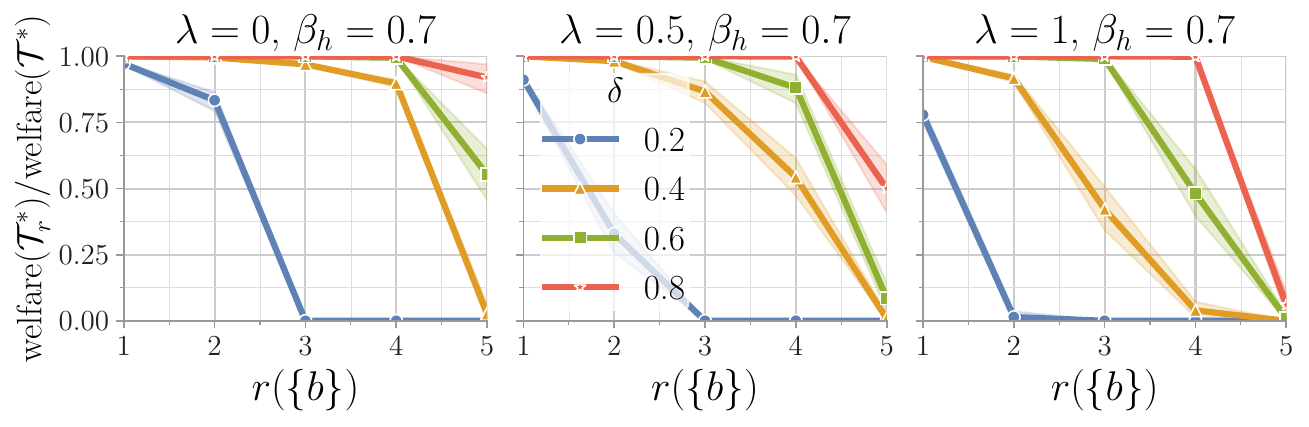}
    \caption{\normalfont The loss in welfare as the lower bound on the number of water
	units for each individual buyer is increased. The welfare-fairness
	tradeoff is the ratio  
    $\welfare(\trade_r^*)$/$\welfare(\trade^*)$,
    where~$\trade_r^*$ is
    an optimal solution which satisfies the
	fairness criteria that every buyer is matched with at
    least~$r(\{b\})=r$ water units and~$\trade^*$ is
    an optimal solution when no such constraints are imposed.
	The results are for the synthetic graphs for different values
	of~$\lambda$ and~$\delta$ over 100 replicates.}
    \label{fig:fairness}
\end{figure}

\para{Buyer satisfaction} For the synthetic networks, we find welfare
maximizing solutions with the constraint that every buyer~$b$ is matched to
at least~$r(\{b\})=r$ water units. Figure~\ref{fig:fairness} shows the
decrease in welfare as~$r$ increases. A value of zero on the y-axis
corresponds to an infeasible instance given the minimum satisfaction
constraints. We note that for lower~$\delta$, the maximum welfare
achievable is small for even small~$r$, indicating that, during water
scarcity, the welfare--fairness trade off is high. For a high~$\lambda$, where
most buyers are low-valued and most sellers are high-valued, we see a sharp
drop in welfare with increasing~$r$.  

\begin{figure}[htb]
    \centering
    \includegraphics[width=\columnwidth]{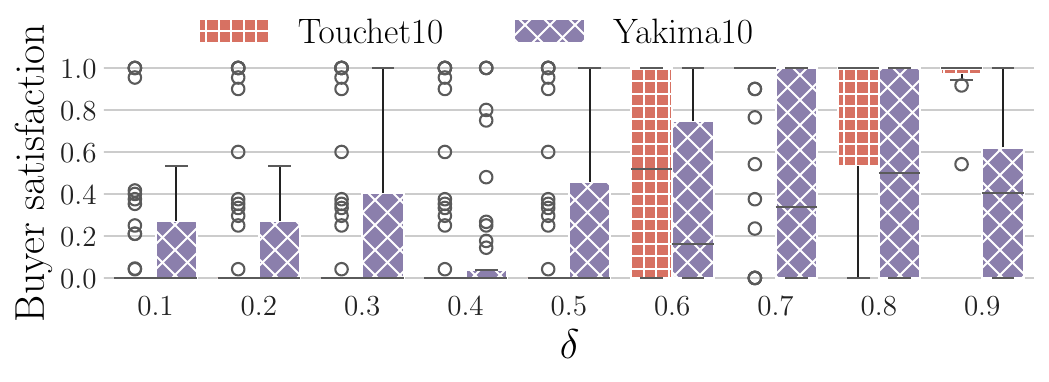}
    \caption{\normalfont A boxplot of buyer satisfaction (water units
    matched/required water units) distribution with respect to water
    availability. Given the optimal
solution from Algorithm~\ref{alg:monotone}, for each buyer, the
satisfaction is computed. In some cases, the boxes are not visible. These
correspond to a median of either zero or one.}
    \label{fig:sat}
\end{figure}

For the real-world graphs, we have plotted buyer satisfaction in
Figure~\ref{fig:sat} for water unit size of 10, when there are no lower
bound constraints. We observe that, for both
datasets, the general satisfaction levels
are low for~$\delta<0.5$.  We see some outliers with 100\% satisfaction.
Upon inspection, we found that these buyers typically have a single unit requirement. We
note that for \touchet{10}, the average buyer satisfaction jumps to
$\approx0.5$ at~$\delta=0.6$. There are low-valued agents with water
available at~$\delta\ge0.5$ who can lead to an increase in trade. We
observe that welfare is not indicative of buyer satisfaction, as it only
depends on the total value and total quantity of trade. We observe that, in
general, it is challenging to guarantee a minimum number of water units to
most buyers due to the fact that, in many scenarios, there are no
compatible sellers for most buyers. Therefore, with the additional
constraints of lower bounds as in Section~\ref{sse:maxwelfarefair}, this
will mostly lead to infeasible instances. Next, we note that the mean or
median buyer satisfaction need not increase as~$\delta$ increases.
As~$\delta$ increases, the number of buyers decreases. Hence, it is
possible that for the remaining  buyers, the buyer satisfaction is low.
This explains the decrease  in buyer satisfaction for \yakima{10}
from~$\delta=0.3$ to~$\delta=0.4$.

\para{The size of water units} We recall that all our proposed algorithms
run in polynomial time in the number of water units. This is unlike the
problems considered in Liu~et~al.~\cite{liu2016optimizing}, where the
complexity was with respect to the number of agents. Given the
heterogeneity in the valuation of each water unit, this is unavoidable.
One way to mitigate this problem is to increase the size of a single water
unit. Our analysis on the size of water units is in the supplement.

\section{Future Work}
\label{sec:concl}
We presented results for a class of resource matching problems motivated by applications to water trading.
One direction for future work is to consider optimal
allocation problems with other welfare functions and fairness criteria.
Our work assumes that the valuations of resources are public. 
If the valuations are not fully revealed (a more
practical setting), interesting and richer 
problems involving negotiations and price
discovery emerge. Our work is a 
 step towards modeling and understanding these issues.

\smallskip 

\clearpage 

\section{Acknowledgments}
{
\small
We thank the reviewers and editors for their valuable suggestions. This
material is based upon work supported by the AI Research Institutes program
supported by USDA-NIFA and NSF under the AI Institute: Agricultural AI for
Transforming Workforce and Decision Support (AgAID) award No.
2021-67021-35344, USDA-NIFA under the Network Models of Food Systems and
their Application to Invasive Species Spread, grant no. 2019-67021-29933, USDA National Institute of Food and Agriculture project \#1016467,
and NSF award No. OAC-1916805 (CINES). Opinions, findings, and conclusions
are those of the authors and do not necessarily reflect the view of the
funding entities.
}
\vspace{-.3cm}

\bibliographystyle{ACM-Reference-Format} 
\bibliography{refs}

\clearpage

\appendix
\onecolumn
\newcommand{\rom}[1]{\uppercase\expandafter{\romannumeral #1\relax}}
\begin{center}
\fbox{{\Large\textbf{Technical Supplement}}}
\end{center}

\bigskip\bigskip

\noindent
\textbf{Paper title:}~ \titlestr{}

\bigskip

\section{Additional Material for Section~\ref{sec:intro}}
\label{app_sec:intro}

\medskip

\subsection{Definitions of Some Combinatorial Problems}
\label{app_sse:prob_def}

\medskip 

This subsection provides formal definitions of some 
problems which are used in various sections of the paper.
These definitions can be found in several standard texts
(e.g., \cite{GJ-1979}). 

\medskip

\noindent 
(a) \textsc{Exact Cover by 3-Sets} (\xthrc{})

\smallskip

\noindent
\underline{\textsf{Instance:}} A universal set
$U = \{u_1, u_2, \ldots, u_t\}$, where $t = 3\ell$
for some integer $\ell$;
a collection $C = \{C_1, C_2, \ldots, C_r\}$,
where each $C_j$ is a subset of $U$ and $|C_j| = 3$,
$1 \leq j \leq r$.

\smallskip

\noindent
\underline{\textsf{Question:}}~ Is there a subcollection
$C'$ of $C$ such that (i) the sets in $C'$ are pairwise
disjoint and (ii) the union of all the sets in $C'$
is equal to $U$?

\medskip

It is well known that \xthrc{} 
is \cnp-complete~\cite{GJ-1979}.
Note that when there is a solution to an \xthrc{} instance,
the collection $C'$ must have exactly $\ell = t/3$ sets
since each set in $C'$ has three elements and the sets must be pairwise disjoint.

\medskip 

\noindent 
(b) \textsc{Maximum Weighted Matching in a Bipartite Graph}

\smallskip 

We recall that a \textbf{matching} $M$ in a graph $G(V,E)$ is a subset of edges such that no two edges of $M$ are 
incident on the same node~\cite{West-2003}.
When there are weights on edges, the weight of a matching $M$
is the sum of the weights of the edges in $M$.
A formal definition
of the maximum weighted matching problem for
bipartite graphs is as follows.

\smallskip 

\noindent 
\underline{\textsf{Instance:}} A bipartite graph $G(S, B, E)$,
a weight $w(e)$ for each edge $e \in E$.

\smallskip

\noindent
\underline{\textsf{Requirement:}} A matching $M$ that has
the maximum weight over all the matchings in $G$.

\medskip 
It is well known that a maximum weighted matching in a bipartite graph $G(S, B, E)$ can be computed in time $O((|S|+|B|)^3)$ (see Table \rom{2} of \cite{duan2014linear} for a summary of the available algorithms and their running times).

\medskip

\newcommand{\bmatch}{\textsc{DCS}}

\noindent 
(c) \textsc{Degree Constrained Subgraph} or 
$b$-\textsc{Matching} (\bmatch) 

\smallskip 

\noindent 
\underline{\textsf{Instance:}} A bipartite graph $G(X, Y, E)$,
nonnegative integers $\lambda(v)$ and $\mu(v)$ for each node
$v \in X \cup Y$ such that $\lambda(v) \leq \mu(v)$.

\smallskip 

\noindent
\underline{\textsf{Requirement:}} Is there a subgraph
$G'(X, Y, E')$ of $G$ such that for each $v \in X \cup Y$,
the degree $d(G',v)$ of $v$ in $G'$ satisfies the
condition $\lambda(v) \leq d(G', v) \leq \mu(v)$?

\medskip 

\noindent
It is known that the \bmatch{} problem can be solved
efficiently using a reduction to the matching problem on
bipartite graphs~\cite{Gabow-1983}.
When a solution exists, a corresponding subgraph can
also be obtained efficiently.

\medskip

\newcommand{\mvc}{\textsc{MinVC}}

\noindent 
(d) \textsc{Minimum Vertex Cover} (\mvc{})

\smallskip

\noindent
\underline{\textsf{Instance:}} An undirected graph $G(V,E)$
and a positive integer $k \leq |V|$.

\smallskip

\noindent
\underline{\textsf{Question:}}~ Is there a vertex cover
$V' \subseteq V$ of size at most $k$ for $G$ (i.e., a subset $V'$ of nodes such that $|V'| \leq k$ and for each edge $\{x,y\} \in E$, at least one of $x$ and $y$ is in  $|V'|$)?

\medskip

It is well known that \mvc{} is \cnp-complete~\cite{GJ-1979}.

\bigskip 

\section{Additional Material for Section~\ref{sec:welfare}}
\label{app_sec:welfare}

\medskip

\subsection{Running time of Algorithm~\ref{alg:monotone}} 
The running time of the algorithm is dominated by the computation of a maximum weighted matching in Step~2. A maximum weighted matching of~$G'$ can be computed in $O((|S'|+|B'|)^3)$ time (see Table \rom{2} of ~\cite{duan2014linear} for the available algorithms and their running times).


\subsection{Statement and Proof of Theorem~\ref{thm:hard_dis_buyer_nonmonotone}}
\label{app_sse:hardness_buyer}

\medskip

\noindent
\textbf{Statement of Theorem~\ref{thm:hard_dis_buyer_nonmonotone}:}
Suppose we are given a set of sellers~$S$, buyers~$B$,
their water units, compatibility
graph~$G(S,B,E)$, and value functions satisfying the following condition:
$\forall s\in S$, $\valfun_s$ is a monotone non-decreasing
function. In this setting, \maxwelfare{} is
\cnp-hard.

\medskip

\noindent
\textbf{Proof:} We use a decision version of
\maxwelfare{} where the input includes an additional integer parameter $\lambda$ and the goal is to determine
whether there is an assignment for which the welfare
function has a value of at least $\lambda$.
Our proof of \cnp-hardness is through a reduction from the  \textsc{Exact Cover by 3-Sets}
(\xthrc{}) problem
(defined in Section~\ref{app_sse:prob_def}).

\medskip

Given an instance $I$ of \xthrc{} consisting of a universal set
$U = \{u_1, u_2, \ldots, u_t\}$, where $t = 3\ell$, and
a collection $C = \{C_1, C_2, \ldots, C_r\}$ of 3-element
subsets of $U$, we construct an instance $I'$ of \maxwelfare{}
as follows.
\begin{enumerate}
\item The set $S = \{s_1, s_2, \ldots s_t\}$ of sellers
is in one-to-one correspondence with the set $U$.
Each seller $s_i$ has only one associated water unit $w^i$,
$1 \leq i \leq t$. (Thus, the set of water units on the
seller side is in one-to-one correspondence with
the set of sellers.)

\item The set $B = \{b_1, b_2, \ldots b_r\}$ of
buyers is in one-to-one correspondence
with the set collection $C$.

\item We assume that the elements of $U$ are ordered as
$\{u_1, u_2, \ldots, u_t\}$ and that the elements 
in each subset $C_j \in C$ are listed in the order in which
they appear in $U$.
For each set $C_j = \{u_{j_1}, u_{j_2}, u_{j_3}\}$ of the \xthrc{}
instance $I$, we create three water units
$w^j_1$, $w^j_2$ and $w^j_3$ for buyer $b_j$, $1 \leq j \leq r$.
Water unit $w^j_p$ is compatible with water unit $w^{j_p}$
(of seller $s_{j_p})$, $1 \leq p \leq 3$.
(For example, suppose $C_j = \{u_3, u_4, u_7\}$. We have
three water units for buyer $b_j$, namely 
$w^j_1$, $w^j_2$ an $w^j_3$; further, 
$w^j_1$, $w^j_2$ an $w^j_3$ are compatible with the seller
water units $w^3$, $w^4$ and $w^7$ respectively.)

\item For each seller $s_i$, the value function $\valfun_{s_i}$
is defined as follows:
$\valfun_{s_i}(w^i) = 1$, $1 \leq i \leq t$.
(Thus, the function $\valfun_s${} associated with 
each seller $s$ is trivially monotone and non-decreasing.)

\item For each buyer $b_j$, the value function $\valfun_{b_j}$
is defined as follows:
$\valfun_{b_j}(w^j_1) = 0$,
$\valfun_{b_j}(w^j_2) = 0$, and
$\valfun_{b_j}(w^j_3) = Q$, where $Q$ 
is an integer $\geq 4$,
$1 \leq j \leq r$.
(The function $\valfun_b${} associated with each buyer $b$  is monotone but non-decreasing; that is, it fails to satisfy the condition needed
for Theorem~\ref{thm:monotone}.)

\item The lower bound $\lambda$  on the value of the welfare
function is set to $\ell(Q-3)$.
\end{enumerate}
This completes the construction of the \maxwelfare{} instance $I'$.
Clearly, the construction can be done
in polynomial time.
We now show that there is a solution to the \maxwelfare{} instance $I'$
iff there is a solution to the \xthrc{} instance $I$.

\medskip

Suppose there is a solution $C'$ to the \xthrc{} instance $I$.
Recall that such a solution must have exactly $\ell$ sets.
Without loss of generality, let the solution be given by
$C' = \{C_1, C_2, \ldots, C_{\ell}\}$.
If $C_j$ = $\{u_{j_1}, u_{j_2}, u_{j_3}\}$,
assign the water units $w^{j_1}$, $w^{j_2}$ and $w^{j_3}$ 
of sellers $s_{j_1}$, $s_{j_2}$ and $s_{j_3}$ respectively
to the three units $w^j_1$, $w^j_2$ and $w^j_3$ of
buyer $b_j$,  $1 \leq j \leq \ell$.
(The other $r-\ell$ buyers are not assigned any water units.)
Since $C'$ covers all the elements of $U$, this assigns all
the $t = 3\ell$ water units of the sellers.
Since the sets in $C'$ are pairwise disjoint, each water unit
of a seller is assigned to exactly one buyer.
Thus, for each seller $s_i$, the value $\valfun_{s_i}(w^i) = 1$
($1 \leq i \leq t = 3\ell$), and
for each buyer $b_j$, the value $\valfun_{b_j}(w^j_3) = Q$.
Hence, the value of the welfare function is
$\sum_{j=1}^{\ell} \valfun_{b_j}(w^J_3) -
\sum_{i=1}^{3\ell} \valfun_{s_i}(w^i)$ = $Q\ell - 3\ell$ =
$\ell(Q-3)$ = $\lambda$.
Thus, we have a solution to the \maxwelfare{} instance $I'$.

\medskip

Now suppose there is a solution to the \maxwelfare{} instance $I'$.
We have the following claim.

\medskip

\noindent
\textbf{Claim 1:} Any valid solution to the \maxwelfare{} 
instance $I'$
must include the following two properties: (a) exactly $\ell$ buyers have
a value of $Q$ for their value functions;
and (b) all the $t = 3\ell$ sellers have the value 1 for
their value functions.

\medskip

\noindent
\textbf{Proof of Claim 1:} First, consider Part~(a).
Since the total number of available water units from the sellers is $t = 3\ell$
and each buyer needs 3 units for their value-functions to have a value $Q$,
at most $\ell$ buyers can have the value $Q$ for their value functions.
Now, suppose for the sake of contradiction, the number $z$ of buyers
with the value $Q$ for their value-functions is \emph{less than} $\ell$.
Such an assignment would use at least $3z$ water units of sellers (since
each buyer with value function $Q$ needs three water units).
Then the value of the welfare function is at most $zQ - 3z$ = $z(Q-3)$.
Since $z < \ell$, the value of the welfare function is
$< \lambda = \ell (Q-3)$. 
This contradicts the assumption that the value
of the welfare function is at least $\ell (Q-3)$, and completes
our proof of Part~(a).

To prove Part~(b), note that from Part~(a), the number of buyers with the
value $Q$ for their value function is $\ell$.
Since each such buyer is assigned three water units, it follows that the total
number of water units assigned to all the buyers is $3\ell$.
This completes the proof of Claim~1. ~$\Box$

\medskip 

In view of Claim~1, any solution to instance $I'$ has exactly $\ell$ buyers,
each of whom has been assigned three water units.
Without loss of generality, let $\{b_1, b_2, \ldots, b_{\ell}\}$
denote this set of buyers.
Construct the following collection $C'$ of subsets: for each buyer $b_j$
in the solution, choose the corresponding set $C_j$ in $C'$,
$1 \leq j \leq \ell$.
Since the sets of water units assigned to the buyers
are pairwise disjoint and the $3\ell$ water units of the sellers
are used in the solution to $I'$, it can be seen that $C'$ is
a solution to the \xthrc{} instance $I'$,
and this completes our proof of
Theorem~\ref{thm:hard_dis_buyer_nonmonotone}. \QED


\bigskip 

\section{Additional Material for 
Section~\ref{sec:fairness}}
\label{app_sec:fairness}

\medskip 

\subsection{Additional Material for 
Section~\ref{sse:maxwelfarefair}}
\label{app_sse:maxwelfarefir}

\medskip 

In Section~\ref{sse:maxwelfarefair}, we considered
the welfare maximization problem when there are lower
bounds on the number of water units to be assigned to
subsets of buyers.
Here, we show that, in general, the problem of 
determining whether there is
a solution that satisfies all such lower bound constraints
is itself \cnp-complete.
It should be noted that this decision problem does not 
involve welfare maximization.
A formal statement of the problem is as follows.

\medskip

\newcommand{\feasdemog}{\textsc{FeasDemog}}

\noindent
\textsc{Feasibility of Demographic Constraints} (\feasdemog)

\smallskip

\noindent
\underline{\textsf{Instance:}}~ A set $S$ of sellers and
their water units, a set $B$ of buyers, a resource--needs
compatibility graph $G'$, a collection $D$
of constraints of the form $(A, q)$ where $A \subseteq B$ 
is a subset of buyers and $q$ is a positive integer 
that gives a lower bound
on the total number of water units to be assigned to the
buyers in $A$.

\smallskip 

\noindent 
\underline{\textsf{Question:}}~ Is there a valid assignment
that satisfies all the constraints in $D$?

\medskip

\noindent
The following result establishes the complexity of the
\feasdemog{} problem.

\medskip

\begin{proposition}\label{pro:feasdemog_hard}
The \feasdemog{} problem is \cnp-complete.
\end{proposition} 

\noindent
\textbf{Proof:} It can be seen that the \feasdemog{} problem
is in \cnp{} since given an assignment, it is easy to efficiently  check whether it satisfies all the constraints in $D$.
To prove \cnp-hardness, we use a reduction from the \mvc{}
problem (defined in Section~\ref{app_sse:prob_def}).

\medskip 

Given an instance $I$ of the \mvc{} problem consisting of a
graph $G(V,E)$ and an integer $k \leq |V|$, we construct an
instance $I'$ of the \feasdemog{} problem as follows.
\begin{enumerate} 
\item The set $S = \{s_1, s_2, \ldots, s_k\}$ consists of $k$
sellers; each seller $s_i$ has exactly one water unit $w_i$,
$1 \leq i \leq k$.

\item Let $|V| = \{v_1, v_2, \ldots, v_n\}$. 
The set $B = \{b_1, b_2, \ldots, b_n\}$ of buyers is in
one-to-one correspondence with the node set $V$.

\item The resource--needs graph $G'$ is a complete bipartite
graph between the water units of sellers and the buyers.
(In other words, any water unit of the sellers can be assigned
to any buyer.)

\item The set $D$ has $m = |E|$ constraints.
For each edge $e = \{v_i, v_j\}$ of $G$, we create a constraint
$(\{b_i, b_j\}, 1)$; that is, the buyers corresponding to the
end points of edge $e$ must be assigned a total of at least
one water unit.
\end{enumerate}
This completes the construction. It is easy to verify that the
construction can be carried out in polynomial time.

\medskip

Suppose there is a solution to the \mvc{} problem.
Without loss of generality, let $V' = \{v_1, v_2, \ldots, v_k\}$
denote the given vertex cover of size $k$.
Consider the matching assignment $\trade$ given by
$\trade$ = $\{(w_i, b_i) ~:~ 1 \leq i \leq k\}$.
(Recall that buyer $b_i$ corresponds to node $v_i$ of $G$.)
Since the resource--needs graph is a complete bipartite graph,
this assignment satisfies the compatibility conditions.
To show that this assignment satisfies all the constraints in $D$,
consider any constraint $(\{b_x, b_y\}, 1)$ in $D$.
This constraint was added to $D$ due to the edge $\{v_x, v_y\}$ in $G$.
Since $V'$ is a vertex cover for $G$, at least one of $v_x$ and $v_y$
appears in $V'$. Without loss of generality, let $v_x$ be in $V'$. 
Thus, the assignment $\trade$ gives a water unit to buyer $b_x$,
thus satisfying the constraint $(\{b_x, b_y\}, 1)$.
Thus, we have a valid solution to the \feasdemog{} instance $I'$.

\medskip 

Now, suppose there is a solution $\trade$  to the \feasdemog{} 
instance $I'$. Without loss of generality, let this solution
assign water units to buyers $\{b_1, b_2, \ldots, b_r\}$ for some
$r \leq k$. (Since there are only $k$ water units in total, the
number $r$ of buyers to whom water units can be assigned 
is at most $k$.)
Let $V' =\{v_1, v_2, \ldots, v_r\}$ be the nodes corresponding
to the buyers to whom water units are assigned by $\trade$.
We now show that $V'$ is a vertex cover for $G$.
Consider any edge $e = \{v_x, v_y\}$ of $G$.
Note that $D$ has the constraint $(\{b_x, b_y\}, 1)$
corresponding to $e$. Since this constraint is satisfied
by $\trade$, at least one of $b_x$ and $b_y$ must be 
assigned a water unit. Thus, at least one of these
two nodes appears in $V'$. Thus, $V'$ is a vertex cover
of size $\leq k$ for $G$.
This completes our proof of Proposition~\ref{pro:feasdemog_hard}. \QED

\bigskip

\subsection{Additional Material for Section~\ref{sse:leximin_fair}}
\label{app_sec:leximin_fair}

\newcommand{\reward}{\mu}
\newcommand{\increward}{\delta}
\newcommand{\totreward}{\mathbb{B}}
\newcommand{\mbsoln}{\mathcal{M}}

\medskip 

The purpose of this section is to show that the \maxleximin{}
problem can be solved efficiently.
For the reader's convenience, we repeat the definition of
the problem.

\medskip 

\noindent 
\textbf{Definition of \maxleximin{} problem:}
Given a seller with a set $W$ of $k$ water units, a set of buyers~$B$ with the same cost for every
water unit, and a compatibility 
graph~$G_w(W,B,E_w)$, find a trading
assignment~$\trade^*$ with the leximin-largest buyer satisfaction vector.

\medskip 

As mentioned in Section~\ref{sse:leximin_fair}, we obtain
an efficient algorithm for the \maxleximin{} problem by reducing
it to a problem called \maxtbmrm{} from Trabelsi~et~al.~\cite{trabelsi2023resource}.
A definition of the \maxtbmrm{} problem is as follows.

\medskip 

\noindent 
\underline{\textsf{Instance:}} A bipartite  graph $G(X, Y, E')$, 
where $X$ is a set of agents and $Y$ is a set of resources, a number of matching rounds, $k$. For each agent~$x_j$, permissible set of rounds~$K_j \subseteq [k]$, the desired number of rounds~$\rho_j$, and a valid benefit
function~$\reward'_j(\cdot)$.

\smallskip

\noindent 
\underline{\textsf{Required}}: Find a collection $\mathcal{M}^*$
consisting of at most $k$ matchings of $G$ that maximizes the sum $\sum_{x_j \in X} \reward'_j(\cdot)$.

\medskip

The efficient algorithm for \maxtbmrm{} presented in \cite{trabelsi2023resource} requires that the benefit functions
for agents be \textbf{valid}. The definition of a valid
benefit function is as follows.

\medskip

\noindent
\textbf{Valid benefit function:}~
For each agent $x_j$, let~$\reward_j(\ell)$ denote the
non-negative \emph{benefit} that the agent~$x_j$ receives if it appears in $\ell$ matchings, for~$\ell=0,1,2,\ldots,\rho_j$.
Let~$\increward_j(\ell)=\reward_j(\ell)-
\reward_j(\ell-1)$ for~$\ell=1,2,\ldots,\rho_j$.
We say that the benefit function $\reward_j$
is \textbf{valid} if it satisfies
\emph{all} of the following four properties:
(P1) $\reward_j(0)=0$;
(P2) $\reward_j$ is monotone non-decreasing in~$\ell$;
(P3) $\reward_j$ has the \emph{diminishing returns property}, that is,
$\reward_j(\ell)-\reward_i(\ell-1)\le\reward_j(\ell-1)-\reward_i(\ell-2)$
for~$2\le\ell\le\rho_i$; and
(P4) $\increward_j(\ell) \leq 1$, for~$\ell=1,2,\ldots,\rho_j$.

\smallskip 

Note that~$\reward_j(\cdot)$ satisfies
property~P3 iff~$\increward_j(\cdot)$ is monotone non-increasing in~$\ell$.

\medskip 

We now show that the following problem, which
is related to the \maxleximin{} problem, can be efficiently
solved through a reduction to the \maxtbmrm{} problem.
(This reduction was sketched in Section~\ref{sse:leximin_fair}.) 

\medskip 

\newcommand{\maxtrade}{\textsc{Max-Reward-Trade}}

\noindent 
\textbf{Finding a Trading Assignment to Maximize the Total Reward}
(\maxtrade)

\smallskip

\noindent
\underline{\textsf{Instance:}} A set $W = \{w_1, w_2, \ldots, w_t\}$
of water units, a set $B = \{b_1, b_2, \ldots, b_m\}$ of buyers,
a bipartite compatibility graph $G(W, B, E)$, a requirement
$\rho_j \leq |W|$ and a valid benefit
function $\reward_j$ for each buyer $b_j$.
(The benefit function $\reward_j(\ell)$ is defined for all $\ell$,
$0 \leq \ell \leq \rho_j$.)

\smallskip

\noindent
\underline{\textsf{Required:}}~ Find a trading assignment $\trade$
such that the total reward 
$\totreward(\trade)$ = $\sum_{j=1}^{m} \reward_j(\cdot)$ due
to the assignment is maximized.

\begin{lemma}\label{lemma:reduction}
The \maxtrade{} problem can be solved in polynomial time
through a reduction to the \maxtbmrm{} problem.
\end{lemma}

\noindent
\textbf{Proof:} A reduction from the \maxtrade{} problem to
the \maxtbmrm{} problem is as follows.
\begin{enumerate}
\item The compatibility graph $G'(X, Y, E')$ 
is a graph with $X=B$, $Y=\{v\}$ is a single node which represents one different water right per matching round and 
$E'=\{\{x,y\}|x\in X,y\in Y\}$.
(Thus, $G'$ is a star graph with a center node and  $|B|$ leaves).
\item  The total number of rounds is $k = |W|$ (i.e., the number of water units).
\item For each buyer $b_j$, $K_j=\{i|(v_i,v_j)\in E\}$
(one round per compatible water unit).
\item For each buyer $b_j$, the desired number of rounds is $\rho_j$. 
\item For each buyer $b_j$, the benefit function is $\reward_j$.
\end{enumerate}

Given a solution~$\mbsoln$ to ~\maxtbmrm{},
we construct a trading assignment $\trade$ as follows:
each matching edge in $\mbsoln$ becomes one entry $(w_i, b_j)$
of $\trade$; here, $w_i$ is determined by the matching round and $b_j$ by the agent.
To see that an optimal solution to \maxtbmrm{} provides an
optimal solution to \maxtrade{}, consider an optimal solution
$\mbsoln^*$ to \maxtbmrm{} and let this correspond to a
trading assignment $\trade$ for \maxtrade{}.
Assume for the sake of contradiction that
there is a solution $\trade'$ such that 
$\totreward(\trade') > \totreward(\trade)$.
Given $\trade'$, we construct a matching solution $\mbsoln'$ such that for each buyer and assigned water right in $\trade'$, we match the corresponding buyer to the water unit in the corresponding round. It can be seen that the resulting solution is valid to \maxtbmrm{}. Furthermore, its value is greater than that of $\totreward(\mbsoln^*)$, contradicting the 
assumption that $\mbsoln^*$ is an optimal solution to the
\maxtbmrm{} instance. To conclude, we have an optimal solution to 
the \maxtrade{} instance.~$\Box$

\medskip 

When the benefit function $\reward_j$ for each buyer
$b_j$ has the form $\reward_j(\ell) = \ell$, for
$0 \leq \ell \leq \rho_j$, the total reward of all the buyers represents the total number of water units assigned. Thus, the above reduction
shows that the problem of finding a trading assignment
that  maximizes the total number
of water units assigned in the above setting can be
solved efficiently.

\medskip 

We now address the main result which is to show that the algorithm for \maxtbmrm{} in \cite{trabelsi2023resource} can be used to solve
the \maxleximin{} problem.
This is done by showing that a valid reward function 
for each buyer
can be constructed so that maximizing the total reward
produces a solution to the \maxleximin{} problem.
We state this result below.

\medskip

\noindent
\begin{lemma}\label{lem:leximin}
Consider any instance of the \maxleximin{} problem.
There exists a valid benefit function $\mu_j$ for each buyer $b_j$ such that maximizing the total benefit under this function gives a solution to the \maxleximin{} problem.
\end{lemma}

We proceed to prove Lemma~\ref{lem:leximin} in several stages.
We first give the procedure for constructing a valid benefit
function for each buyer.  The steps of this procedure are
as follows. (The reader should bear in mind that $B$ and $W$
denote the set of buyers and water units, respectively, of
the \maxleximin{} problem. Also, for each buyer $b_j$,
$\rho_j$ gives the number of water units desired by $b_j$.)

\newcommand{\Fg}{F^>}
\newcommand{\ord}{\pi}

\smallskip

\noindent
\begin{enumerate}
\item Let~$F=\big\{k_i/k_j\mid k_i,k_j\in [W]\} ~\mathrm{and}~ k_i \leq k_j\big\}\cup\{0\}$.
\item Let~$\ord: F\rightarrow\{1,\ldots,|F|\}$ correspond to the index of
each element in~$F$ when sorted in descending order.
\item For each~$q\in F$, let $\xi(q)=(|B|\cdot |W|)^{\ord(q)}/(|B|\cdot |W|)^{|F|}$. Note that
each~$\xi(q)\in(0,1]$.
\item The incremental benefit~$\delta_j(\ell)$ for a buyer~$b_j$ for
the~$\ell$th water unit is defined
as~$\delta_j(\ell)=\xi\big((\ell-1)/\rho_j)\big)$.
Thus, the benefit function $\mu_j$ for buyer $v_j$ is
given by $\mu_j(0) = 0$ and
$\mu_j(\ell) = \mu_j(\ell-1) + \delta_j(\ell)$ for
$1 \leq \ell \leq \rho_j$.
\end{enumerate}

\medskip 

It can be seen that $\mu_j$ satisfies
properties P1, P2 and P4 of a valid benefit
function. We now show that it also satisfies P3,
the diminishing returns property.

\begin{lemma}\label{lem:dim_returns}
For each buyer~$b_j$, the incremental benefit~$\delta_j(\ell)$ is monotone
non-increasing in~$\ell$. Therefore,~$\mu_j(\cdot)$ satisfies
diminishing returns property.
\end{lemma}
\begin{proof}
By definition, for~$\ell=1,\ldots,\rho_j-1$,~$\delta_j(\ell)/\delta_j(\ell+1)=
\xi\big((\ell-1)/\rho_j\big)/\xi\big(\ell/\rho_j\big)\ge |B|\cdot|W| > 1$, by noting
that~$\pi(\ell/\rho_j)-\pi((\ell+1)/\rho_j)\ge1$. 
Hence, the lemma holds.
\end{proof}

To complete the proof of Lemma~\ref{lem:leximin}, we must show that any solution that maximizes the benefit function defined above optimizes the leximin order.
We start with some definitions and a lemma.

\smallskip

For a buyer~$b_j$ and~$q\in F$, let~$\Fg_j(q)=\{\ell\mid 0\le\ell\le\rho_j~
\text{ and } \ell/\rho_j >q\}$. We will now show that the incremental
benefit~$\delta_j(\ell)$ obtained by buyer~$b_j$ for matching~$\ell$ is
greater than the sum of all incremental benefits~$\delta_{j'}(\ell')$, for
all buyers~$b_{j'}$ and for all~$\ell'$
satisfying~$\ell'/\rho_{j'})>\ell/\rho_j$.
\begin{lemma}\label{lem:cost}
For any buyer~$b_j$ and non-negative
integer~$\ell\le\rho_j$, we have ~$\delta_j(\ell)>\sum_{j'\in B}\sum_{\ell'\in
\Fg_{j'}(\ell/\rho_j)}\delta_{j'}(\ell')$.
\end{lemma}

\begin{proof}
For any~$\ell'\in \Fg_{j'}(\ell/\rho_j)$, note that
$\ell/\rho_j<\ell'/\rho_{j'}$.
Hence,~$\pi(\ell/\rho_j)-\pi(\ell'/\rho_{j'})\ge 1$. 
This implies
that~$\delta_j(\ell)/\delta_{j'}(\ell') =
\xi\big((\ell-1)/\rho_j\big)/\xi\big((\ell'-1)/\rho_{j'})\big)\ge |W|\cdot|B|$.
Since~$\rho{j'}\le |W|$, the number of
$\delta_{j'}(\ell')$ terms
per~$j'$ is at most $|W|$. 
Since $\ell/\rho_j \notin \Fg_{j'}(\ell/\rho_j)$, the number of $\delta_{j'}(\ell)$ terms for $j'=i$ is at most $|W|-1$. Since there are~$|B|$ buyers~$b_{j'}\in B$, it
follows that there are at most~$|B|\cdot|W|-1$ 
terms~$\delta_{j'}(\ell')$ in total. Therefore,~$\sum_{j'=i}\sum_{\ell'\in
\Fg_{j'}(\ell/\rho_j}\frac{\delta_{j'}(\ell')}{\delta_{j}(\ell)}<1$.
\end{proof}

\smallskip
We are now ready to complete the proof of Lemma~\ref{lem:leximin}.

\medskip


\noindent
\textbf{Proof of Lemma~\ref{lem:leximin}.}
The proof is by contradiction. Suppose~$\mathcal{A}^*$ is an optimal solution
given the benefit function defined above. We will show that if there exists
a solution~$\mathcal{A}$ with leximin order better than that
of~$\mathcal{A}^*$, then the total
benefit~$\totreward(\mathcal{A})>\totreward(\mathcal{A}^*)$, contradicting the fact
that~$\mathcal{A}^*$ is an optimal solution.
Let $A=(l_1/\rho_1,l_2/\rho_2,..., l_{|B|}/\rho_{|B|})$ be a sorted list of satisfaction ratios for $\mathcal{A}$ and let $A^*=(l^*_1/\rho_1,l^*_2/\rho_2,..., l^*_{|B|}/\rho_{|B|})$ be a list defined similarly for $\mathcal{A}^*$.
Let~$k$ denote the first index for which $l_k/\rho_k \neq l^*_k/\rho_k$.

\medskip 

For a given solution $\mathcal{A}$, we use $\gamma_j(\mathcal{A})$
to denote the number of water rights assigned to buyer $b_j$,
$1 \leq j \leq |B|$.
We recall that the benefit function for~$\mathcal{A}$ can be written
as~$\totreward(\mathcal{A})=\sum_{j=1}^{|B|}\sum_{\ell=1}^{\gamma_j(\mathcal{A})}
\delta_j(\ell)$. 
For each~$j$, the~$\delta_j(\ell)$ terms can be
partitioned into two blocks as follows.

\begin{description}
\item{(i)} $D_1(\mathcal{A})=\{(j,\ell)\mid \forall
j,~\ell/\rho_j <k\}$,  and
\item{(ii)} $D_2(\mathcal{A})=\{(j,\ell)\mid
\forall j,~\ell/\rho_j\geq k$.
\end{description}

\noindent
We can
partition the terms corresponding to~$\totreward(\mathcal{A}^*)$ in the same
way, and these blocks are denoted by~$D_1(\mathcal{A}^*)$ and $D_2(\mathcal{A}^*)$.
From the definition of $k$, it follows that~$D_1(\mathcal{A}^*)=D_1(\mathcal{A})$. Since $\mathcal{A}$ has a better leximin order, and, from definition, it follows that $\ell^*_k/\rho_k < \ell_k/\rho_k$. In addition, from the order of the ratio lists, it follows that $\ell_k/\rho_k>\ell^*_j/\rho_j$ for all $j>k$.
From lemma~\ref{lem:cost}, it follows that $\delta_k(\ell_k)>\sum_{j'\in B}\sum_{\ell'\in
\Fg_{j'}(\ell/\rho_j}\delta_{j'}(\ell')$, and, therefore, $\delta_k(\ell_k)>\sum_{(j,\ell)\in D_2(\mathcal{A}^*)}\delta_j(\ell)$.
Finally, we get \begin{align*}
\totreward(\mathcal{A})-\totreward(\mathcal{A}^*) = &
\sum_{(j,\ell)\in D_2(\mathcal{A})}\delta_j(\ell) -
\sum_{(j,\ell)\in D_2(\mathcal{A}^*)}\delta_j(\ell) \\
& \geq
\delta_k(\ell_k) -
\sum_{(j,\ell)\in D_2(\mathcal{A}^*)}\delta_j(\ell) > 0\\
\end{align*}

Thus, $\totreward(\mathcal{A})>\totreward(\mathcal{A}^*)$, and this
contradicts the optimality of $\mathcal{A}^*$.
The lemma follows. \qed

\medskip 

Thus, we have shown that any given instance of the \maxleximin{} problem can be solved efficiently by
constructing suitable benefit functions for each buyer,
reducing the problem to the \maxtbmrm{} problem, and
then using the efficient algorithm for the \maxtbmrm{}
problem from \cite{trabelsi2023resource}.
We have thus completed a proof of the following
theorem (i.e.,
Theorem~\ref{thm:leximin} of 
Section~\ref{sse:leximin_fair}):

\medskip

\noindent
\textbf{Theorem~\ref{thm:leximin}:} An optimal solution to the \maxleximin{} problem can be obtained in polynomial time.

\bigskip 

\section{Additional Material for Section~\ref{sec:exp}}
\label{app_sec:rw-dataset}

\begin{table}[htb]
    \caption{Datasets and their attributes.}
    \small
\centering
\begin{tabular}{lp{.7cm}p{1.0cm}p{1.3cm}p{1.0cm}p{1.0cm}}
\toprule
Data &
Num. agents &
Total water units &
Total value &
Max. value &
Min. value \\
\midrule
Synthetic &
10 &
50 &
variable &
variable &
variable \\ 
\touchet{5} &
93 &
2704 &
8,533,475.94 &
57629.24 &
703.38 \\ 
\touchet{10} &
93 &
1389 &
8,778,037.64 &
115258.49 &
1406.77 \\ 
\touchet{20} &
93 &
741 &
9,400,712.72 &
230516.97 &
2813.54 \\ 
\yakima{5} &
77 &
3616 &
21,838,629.76 &
31606.52 &
264.33 \\ 
\yakima{10} &
77 &
1838 &
22,088,453.49 &
63213.04 &
528.66 \\ 
\yakima{20} &
77 &
952 &
22,698,607.80 &
126426.07 &
1057.31 \\ 
\bottomrule
\end{tabular}

    \label{tab:data}
\end{table}
The real-world networks were constructed using water rights data, acreage, types
of crops grown, and water demand by crop
type~\cite{waterrights,aglanduse,nass2022}.
We identified 93 water rights held in the
Touchet River Watershed and 77 water rights held in the Yakima River Basin.
Both of these rivers are in Washington State. The water rights were correlated with their
associated farm fields, which allowed the water rights to be mapped to
their corresponding crop types, acreage, and water demand. Using Washington
State crop production data from the USDA National Agricultural Statistics
Service (NASS)~\cite{nass2022}, we estimated the value of production per
acre for each crop type, and made educated guesses on 
those crop types which were not included in NASS data.
We then calculated the volume of water needed per water right crop
type, and assigned a monetary value to each unit of water needed.
To assign the value per unit, we first calculated the total volume $v_i$
needed per field $i$ in acre-feet as $v_i$ = round($a_i\cdot mm_i/304.8$),
where $a_i$ and $mm_i$ denote respectively the acreage and the volume required per acre (in acre-millimeters)
of field $i$.

The total (baseline) water capacity $W$ was calculated as the sum of demands
across all of the water rights per watershed (Touchet and Yakima) under
normal conditions. The reduced water capacity $R$ -- to account for drought
conditions -- was calculated as $R_i$ = $c_i$/100 * $W$ for each
percentage capacity $c$ in [10\% .. 90\%], calculated in 10\% increments.
Sorting the water rights by seniority (based on priority date), more senior
water rights were designated as \emph{sellers} until the sum of the
sellers' water volumes would have exceeded $R$. All other, more junior,
water rights were designated as \emph{buyers}. Then, for each seller and
buyer, we disaggregated the total volume and value per unit into separate
rows by unit. For the seller, it was assumed that they would prioritize
selling water for their lower value crops first, so the array of
disaggregated units would be sorted by value per unit in ascending order.
Similarly, buyers would probably prioritize buying water for their
high-value crops first, so their array of disaggregated units was sorted by
value per unit in descending order.

Finally, a bipartite graph was created, mapping each of the sellers' units of water to compatible buyers' units. In order to be compatible, ($i$) the buyer had to be upstream or downstream of the seller (e.g., not in different streamsheds or in different forks of the same river), and ($ii$) the buyer's value for a given unit had to be greater than or equal to the seller's value (or price) for that unit. 

A summary of values for Yakima -- based on different $R$ values -- can be seen in Table \ref{table:rw-dataset}. A few observations are worth noting. The Unit Size in Acre-Feet indicates that unit size was calculated between rows for 10 and 20 acre-feet per unit, which explains why the 20 acre-feet number of units will always be close to 50\% of the 10 acre-feet number of units. Second, when dividing the acre-feet per unit size, any partial units are rounded up to the next integer value, which is why the Total Value for the 20 acre-feet unit size will always be somewhat larger than the Total Value for the corresponding 10 acre-feet unit size. There appears to be no difference between the 10\% water capacity and 20\% water capacity statistics because the second most senior water right included more than 20\% of the total water units, which prevented the number of sellers to increase until total water capacity rose to 30\%.  Finally, these values are based only on the statistics for the dataset pre-experiment, or the number of units and total value that could possibly be traded, but not the actual totals from matching done during trading.

\begin{table}[h!]
\centering
\begin{tabular}{||c c c c c c||} 
 \hline
 \% Total Water Capacity & Unit Size in Acre-Feet & \# Units & Total Value & Sellers & Buyers \\ [0.5ex] 
 \hline\hline
 10\% & 10 & 11 & \$27534.75 & 1 & 76\\ 
 10\% & 20 & 6 & \$30037.91 & 1 & 76 \\
 20\% & 10 & 11 & \$27534.75 & 1 & 76\\ 
 20\% & 20 & 6 & \$30037.91 & 1 & 76 \\ 
 30\% & 10 & 542 & \$10,664,614.78 & 8 & 69\\ 
 30\% & 20 & 276 & \$10,795,007.24 & 8 & 69 \\
 40\% & 10 & 705 & \$14,590,442.54 & 26 & 51\\ 
 40\% & 20 & 366 & \$14,872,635.68 & 26 & 51 \\
 50\% & 10 & 905 & \$14,980,581.61 & 37 & 40\\ 
 50\% & 20 & 471 & \$15,287,062.66 & 37 & 40 \\
 60\% & 10 & 1060 & \$15,321,802.09 & 46 & 31\\ 
 60\% & 20 & 551 & \$15,639,329.06 & 46 & 31 \\
 70\% & 10 & 1270 & \$19,918,785.39 & 57 & 20\\ 
 70\% & 20 & 659 & \$20,363,049.477 & 57 & 20 \\
 80\% & 10 & 1435 & \$20,386,759.85 & 63 & 14\\ 
 80\% & 20 & 745 & \$20,882,922.60 & 63 & 14 \\
 90\% & 10 & 1556 & \$20,630,547.56 & 68 & 9\\ 
 90\% & 20 & 807 & \$21,134,945.99 & 68 & 9 \\
  [1ex] 
 \hline
\end{tabular}
\caption{Value breakdown and seller-buyer ratios for the Yakima dataset for unit-sizes 10 and 20 acre-feet per unit.}
\label{table:rw-dataset}
\end{table}

\para{The size of water units} We recall that all our proposed algorithms
run in polynomial time in the number of water units. This is unlike the
problems considered in Liu~et~al.~\cite{liu2016optimizing}, where the
complexity was with respect to the number of agents. Given the
heterogeneity in the valuation of each water unit, this is unavoidable.
One way to mitigate this problem is to increase the size of a single water
unit. 
Our results in Figure~\ref{fig:real} compare networks with different water unit
resolutions with respect to both the structural properties and the welfare
from trade. We note that Plot~(a) is not affected by the coarsening of the
resolution, i.e., the number of sellers remains the same as a function
of~$\delta$. However, we do see a benefit of coarsening, as it greatly
reduces the number of edges in the resources--needs compatibility graphs as
the size of a water unit is increased. However, we observe that coarsening
has very little effect on total value. We do observe a significant
difference in the welfare curves of \touchet{10} and \touchet{20}. We note
that for any fixed value of~$\delta$, the greater the size of the water
unit, the greater is the welfare. This is mainly due to rounding effects
that lead to an increase in the total number of water units. However, the
trend remains the same across different sizes.


\end{document}